\documentclass[oldversion]{aa}

\def\cd#1{\textbf{#1}}
\def\cd#1{#1}
\def\um{\ensuremath{\mu\mbox{m}}\xspace}
\usepackage{amsmath,amsbsy}
\usepackage{xspace}
\usepackage[bf,small]{caption}
\usepackage[varg]{txfonts}
\usepackage{setspace}
\usepackage{longtable}
\usepackage[usenames]{color}

\usepackage{graphicx}
\usepackage{natbib}

\newcommand{\psig}{\phi_\sigma}
\newcommand{\tpsig}{$\phi_\sigma$}
\newcommand{\mr}[1]{\mathrm{#1}}
\newcommand{\ep}[1]{\cdot 10^{#1}}

\definecolor{orange}{rgb}{1.0,0.3,0} 
\definecolor{darkblue}{rgb}{0,0,0.8} 
\definecolor{darkgreen}{rgb}{0,0.6,0} 
\definecolor{purple}{rgb}{0.5,0,0.5}

\newcommand{\se}[1]{\mbox{Sect.\ \ref{sec:#1}}}
\newcommand{\Se}[1]{\mbox{Section\ \ref{sec:#1}}}
\newcommand{\eq}[1]{\mbox{Eq.\ (\ref{eq:#1})}}

\newcommand{\fg}[1]{\mbox{Fig.\ \ref{fig:#1}}}
\newcommand{\Fg}[1]{\mbox{Figure\ \ref{fig:#1}}}
\newcommand{\Tb}[1]{\mbox{Table\ \ref{tab:#1}}}
\newcommand{\tb}[1]{\mbox{Tab.\ \ref{tab:#1}}}

\newcommand{\labelH}[1]{\label{#1}} 

\newcommand{\ie}{i.e.,}
\newcommand{\eg}{e.g.,}

\newcommand{\figlarge}[3]{
\begin{figure*}
\centering
\includegraphics[width=15cm]{#1}
\caption{#2}
\label{fig:#3}
\end{figure*}
}

\newcommand{\figsmall}[3]{
\begin{figure}
\resizebox{\hsize}{!}{\includegraphics{#1}}
\caption{#2}
\label{fig:#3}
\end{figure}
}

\begin{document}
   \title{Collisional evolution of dust aggregates. 
     From compaction to catastrophic destruction.}

\titlerunning{Collisional evolution of dust aggregates}
\authorrunning{D. Paszun \& C. Dominik}

\author{D. Paszun$^{1}$ \& C.~Dominik$^{1,2}$}
\institute{$^{1}$Sterrenkundig Instituut `Anton Pannekoek', Kruislaan 403,
NL-1098 SJ Amsterdam, The Netherlands; e--mail:
dominik@science.uva.nl\\
$^{2}$Afdeling Sterrenkunde, Radboud Universiteit Nijmegen,
Postbus 9010, 6500 GL Nijmegen}

\offprints{C. Dominik, \email{dominik@uva.nl}}

\date{Received $<$date$>$ / Accepted $<$date$>$}

\abstract{ The coagulation of dust aggregates occurs in various
  astrophysical environments. Each one is characterized by different
  conditions that influence the growth, \eg relative velocities,
  composition, and size of the smallest constituents (monomers). Here
  we study the microphysics of collisions of dust aggregates in a
  four-dimensional parameter space.  The parameters are the collision
  energy, the initial compactness of agglomerates, the mass ratio of
  collision partners, and the impact parameter.  For this purpose we
  employ a state of the art \emph{molecular dynamics} type of model
  that has been extensively and successfully tested against laboratory
  experiments. It simulates the motion of individual monomers
  interacting dynamically via van der Waals surface forces. The
  \cd{structure of aggregates is quantified by the filling factor that
    provides information about the internal structure, the packing
    density of monomers, and the projected surface area of
    aggregates.}

  Our results show the importance of the impact parameter that causes
  formation of elongated particles, due to tensile forces acting in
  offset collisions.  In head-on impacts, aggregates are compacted at
  lower energies.  A sufficiently high energy causes restructuring to
  reach maximum compaction.  If more energy is provided, pancake-like
  structures are formed.  We find that the outcome of collisions can
  be represented in a simple way. A highly pronounced large fragment
  component coexists with a power-law distribution of small fragments.
  The structural parameter of these small fragments is described very
  well by a simple relation, largely independent of the initial
  compactness, impact energy or impact parameter.  The simulations
  show that erosion by collisions with high mass-ratio can be
  significant.  The ejected mass can be several orders of magnitude
  higher than the impactor mass.  This contrasts with collisions of
  equal mass aggregates, where the same impact \cd{energy} can lead to
  perfect sticking.

  These findings are summarized in the form of a simple collision
  recipe.  The recipe specifies the outcome of a collision, averaged
  over the impact parameter.  It is provided in tabular form for a
  range of physical parameters such as impact energy and pre-collision
  filling factor.  The dependence on the mass ratio of impactor and
  target is taken into account by providing both a local and a global
  branch of the recipe.}


\maketitle

\section{\labelH{sec:introduction}Introduction}

Dust plays an important role in many astrophysical environments,
\cd{and it is also the building material of planets.}  Small grains
can stick to each other thanks to van der Waals forces
\citep{1971ProcRSocLonA..324.301J,1975JournCollInterfSci...53..314D}. Depending
on the growth mechanism, the aggregates can acquire different
structures.  In low velocity collisions, which are dominant in the
regime where particle velocities are \cd{dominated} by Brownian
motion, the narrow size distribution primarily leads to collisions
between particles of a similar size, in turn leading to fractal
aggregates.
\citep{1999Icar..141..388K,2000PhRvL..85.2426B,2004PhRvL..93b1103K,2006Icar..182..274P}.
Fractals produced by Brownian growth have a typical fractal dimension
of $D_\mathrm{f}=1.5$ \citep{2004PhRvL..93b1103K,2006Icar..182..274P},
and higher gas densities shorten the mean free path of particles and
\cd{result} in even more fluffy structures.  In the limiting case
\cd{of very high densities}, aggregates formed this way may be very
elongated with the fractal dimension approaching unity
\citep{2006Icar..182..274P}.

Velocities induced by turbulent gas motions and by radial drift in
protoplanetary disks can be much higher.  Furthermore, the dependence
\cd{of} velocities on particle size can become reversed, with larger
particles moving faster than smaller particles.  For a discussion of
the relative velocities in protoplanetary disks, we refer the reader to
a review article by \citet{2000prpl.conf..533B}.  \cd{Relative motion
  that emphasizes collisions between particles of \emph{different}
  sizes leads to the production of aggregates with very different
  structures, because then the particles also grow by collecting
  smaller projectiles.  In this case, the results are porous,
  non-fractal aggregates
  \citep{1984PhRvA..29.2966B,2008PaszunHierarchical_inprep}.}

\cd{Low-impact velocities} generally result in sticking.  However, the
growth of dust usually causes an increase in collision velocities when
the particle start to decouple from the surrounding gas.  When the
collision energy becomes higher than the energy needed to roll
monomers over each other (later referred to as rolling energy
$E_\mathrm{roll}$), restructuring begins.  Very fluffy and fractal
aggregates are compacted upon collision
\citep{1997ApJ...480..647D,2000Icar..143..138B}.  A further increase
in the impact velocity leads to compaction.  However, eventually the
kinetic energy is high enough to break contacts between individual
monomers.  Erosion then starts to remove parts of colliding
aggregates.  As the relative velocity increases further, erosion also
becomes stronger, ultimately leading to the destruction of the
aggregates.

Fragmentation of aggregates is a major \cd{obstacle} in planet
formation theory.  Dust cannot grow all the way to form planetesimals
because it gets destroyed once relative velocities become violent
enough to disrupt aggregates.  Similarly, radial drift may prevent
growth by removing particles from the disk once they grow to a certain
size and spiral towards the central star.

\citet{2006ApJ...636.1121J} show that, in the presence of turbulence,
planetesimals might be produced by gravitational collapse of clumps of
meter sized boulders.  These clumps are generated by high-pressure
turbulent eddies that trap and concentrate particles, which then become
gravitationally bound.  Before this process can take place, particles
must already have grown by 18 orders of magnitude in mass. The only
feasible way to do that is by collisional sticking. Because
collisional fragmentation of aggregates may prevent growth of large,
meter sized aggregates, it is crucial to fully understand mechanisms
involved in collisions of porous aggregates.

Recently, \citet{2008A&A...487L...1B} have studied the growth of dust
particles in a local density maximum caused by the evaporation front at
the snow line.  This local pressure maximum in the disk can accumulate
dust particles and reduce relative velocities considerably.  In the
midplane of the disk, weaker turbulence \citep{2007ApJ...654L.159C}
results in lower relative velocities. \citet{2008A&A...487L...1B} have
shown that for the fragmentation threshold velocity of at least 5 m/s,
the growth can proceed to large boulders of up to several 100 meters
in size.  However, a more realistic value of the threshold velocity,
for aggregates made of micron-sized grains, is $\sim 1$ m/s
\citep{1993Icar..106..151B,2000ApJ...533..454P,2008ApJ...675..764L}.

\citet{2000Icar..143..138B} performed laboratory experiments of
collisions of dust aggregates.  They studied impacts at a wide range
of energies, from low energy (perfect sticking), through restructuring
to fragmentation of microscopic aggregates.  Their results with
respect to restructuring threshold are in agreement with theoretical
findings of \citet{1997ApJ...480..647D}.  The fragmentation energy,
however, differs, which is \cd{a consequence of the discrepancy} in
sticking velocity. \citet{2000ApJ...533..454P} \cd{measured the sticking
velocity of micron--sized silica grains and found} $1.2$
m/s. \citet{1993ApJ...407..806C} and \citet{1997ApJ...480..647D} on
the other hand derived theoretically a much lower velocity,
inconsistent with experiments.  \cd{This problem was addressed by
  \citet{2008A&A...484..859P}, who included an additional energy
  dissipation channel in order to match the experimental results.  For
  details see section~\ref{sec:model}.}

\citet{1993Icar..106..151B} studied collisions of macroscopic,
mm--sized, aggregates at velocities between 1 and a few m/s. In this
case, however, a different behavior was observed.  Although particles
were very porous (up to $4\%$ filling factor) they did not observe
restructuring.  Instead, aggregates were bouncing off each other or,
for faster impacts, fragmenting.

Very energetic collisions between large mm--sized and cm-sized
aggregates were studied by \citet{2005Icar..178..253W}.  They showed
that the fragmentation observed at high velocity impacts turns into a
net growth of $50\%$ at velocities above 13 m/s.  The distribution of
fragments at velocities of about 20 m/s followed a power-law with a
slope of $-5.6\pm0.2$ for larger fragments and was flat for the
smallest ejecta.

\citet{1977Icar...31..277F} experimented with solid basalt rocks. High
velocity impacts of a few km/s result in a power-law
distribution of small fragments $n(m) \sim m^k$. The slope of the
distribution was found to be $k=-1.83$. They distinguished several
collisional outputs, depending on the target size:
\begin{enumerate}
\item{complete destruction;}
\item{remaining core;}
\item{transition from core to cratering;}
\item{crater formation.}
\end{enumerate}

\citet{2004Icar..167..431S} has developed a smoothed particles
hydrodynamics (SPH) model to simulate meter sized and larger
aggregates. A single particle in this model corresponds to a porous
material described by compressive strength, tensile strength, density,
and sound speed. This method was also used by
\citet{2007A&A...470..733S}.

Although collisions of dust particles have been studied experimentally
and theoretically, no one has formulated a \cd{quantitative recipe
describing both mass distribution and structural properties of the
collisional output, using a model based on empirical results}. Both the
distribution of masses and the compactness of fragments is required to
fully understand the growth of dust. Here we present an extensive
parameter study of many collisions of small dust aggregates. We
provide a recipe for mass distribution and compactness of fragments.

In \se{model} we briefly present the model we adopt to simulate
collisions of dust aggregates, it's strengths and limitations.  In
\se{results} we present findings of our study and discuss them
qualitatively.  \Se{recipe} provides quantitative description of our
results in a form of a collision recipe.  We end this paper with
interesting conclusions in \se{conclusions}.

%
%
%
%
\section{\labelH{sec:model}The Model}

The simulations presented in this work are done using the N-body
dynamics code SAND\footnote{Soft Aggregate Numerical Dynamics}.  Our
model treats all monomers (also referred to as grains or particles) in
the agglomerates (also referred to as aggregates, clusters, or
particles) individually.  Since we are currently not interested in
long range interactions, electrostatic, magnetic and gravitational
forces are not included even though the code can handle them
\citep{2002Icar..157..173D}.  We calculate motion of individual
monomers that interact with each other via attractive van der Waals
surface forces \citep{1971ProcRSocLonA..324.301J}.

The presence of the attractive surface \cd{on deformable particles}
inevitably leads to several energy dissipation mechanisms.  The
particles, when in contact, may roll over each other.  This rolling
motion is \cd{countered} by a rolling friction force
\citep{1995PhMA...72..783D}, causing \cd{an} energy loss.  The same
happens in the case of a sliding motion.  The contacts may shift which
again is work done against the sliding friction force
\citep{1996PhMA...73.1279D}.  Beside that, every time a contact
between two monomers is broken, the elastic energy stored in it is
partially lost
\citep{1971ProcRSocLonA..324.301J,1993ApJ...407..806C,1997ApJ...480..647D}. Some
energy may also be lost due to a twisting motion of particles in
contact \citep{1997ApJ...480..647D}.

We also include an additional energy dissipation channel in order to
fit the experimental results by \citet{2000ApJ...533..454P}. This
process increases the sticking velocity from about
$10\,\mathrm{cm}\,\mathrm{s}^{-1}$
\citep{1993ApJ...407..806C,1997ApJ...480..647D} to about
$1\,\mr{m}\,\mr{s}^{-1}$ \citep{2000ApJ...533..454P}.  Since the
measured attractive force agrees well with the theory provided by
\citet{1971ProcRSocLonA..324.301J} and
\citet{1975JournCollInterfSci...53..314D}, the difference in sticking
velocity points to additional energy losses in collisions.  While our
model does not specify what this mechanism is, a candidate would be
plastic deformation of surface asperities on nm scales.  In order to
achieve agreement with the experimental results, a mechanism
dissipating the energy upon the first contact of two particles was
introduced \citep{2008A&A...484..859P}.  For details regarding the
implementation, we refer the reader to \citet{2002Icar..157..173D} and
\citet{2008A&A...484..859P}.

Although \citet{2008A&A...484..859P} tested the model extensively
against the laboratory experiments and found a good agreement, this
model has limitations as discussed below.
%
%
%
%
\subsection{\labelH{sec:limitations}Limitations}
Although our model presents a new approach to study collision dynamics of
aggregates, the following limitations apply.
\begin{itemize}
\item{Aggregate size}\\ Two monomers in contact oscillate in relative
  distance due to the competition \cd{between attractive forces
    that} hold them together, and the elastic force that pushes them
  apart.  For micron-sized grains (the monomers we consider here have
  radii of $0.5$\um) the vibration frequency is on the order of GHz.
  A correct simulation of this dynamical system requires that we must
  resolve the shortest timescales.  These very short simulation time
  steps limit the number of monomers in the system we can model. The
  largest aggregates we can simulate are made of $10^5$ grains and
  take several hundreds of hours of a CPU time.  In this study we
  model aggregates made of up to 1000 monomers.

  The collisional outcome depends \cd{foremost} on the strength of
  \cd{individual contacts in the} impacting aggregates, thus the
  properties of monomers (size, composition).  The aggregate size
  determines the amount of energy that can be dissipated ($E_\mr{br}$
  times the number of monomers).  Note, however, that laboratory
  experiments of much larger aggregates (about 100 micron in size),
  made of billions of monomers, show phenomena not seen in the
  smallest aggregates.  An example is bouncing
  \citep{2008ApJ...675..764L}.  Although this is an important effect
  that significantly affects the growth of dust aggregates, it is
  still poorly understood.

\item{Irregular grains}\\ Our model assumes spherical monomers, as the
  simulation of irregular grains is \cd{computationally very
    expensive} and, therefore, is impractical.  Moreover, we can
  directly compare our findings to the results of laboratory
  experiments \citep{2000Icar..143..138B}.  Here we briefly introduce
  the possible effect irregular grains may have on the collision
  outcome.

  The strength of an aggregate strongly depends on the size of the
  contact between monomers.  In the case of randomly shaped grains,
  contact is established between surface asperities and is very much
  limited. This reduces the strength of an aggregate and leads to
  disruption at lower impact energies.  Moreover, irregular monomers
  can form more than one contact with each other, which means that
  individual contact points must break before restructuring can occur.
  In this way, chains of irregular particles can be more rigid than
  those made of spherical particles.

  \citet{2000ApJ...533..454P} performed laboratory experiments on both
  spherical and irregular grains. They showed that irregular particles
  can stick to a flat surface at much higher velocities than spherical
  grains (1.5 to 2.3 m s$^{-1}$ for spherical silica grains and 5 to
  25 m s$^{-1}$ for irregular enstatite monomers). This suggests that
  additional energy dissipation may occur during a collision.
  \cd{The} mechanism, however, is still unclear.

  As the physical processes involved in collisions of irregular grains
  are unknown, we leave them for a future investigation.  However,
  experiments have shown that the overall effects off collisions are
  well reproduced between experiments with spherical or round monomers
  \citep{2008ApJ...675..764L}.  \cd{Therefore, we believe} that the
  general conclusions of our study here will also hold for less
  perfect monomers.

\end{itemize}
%
%
%
%
\subsection{\labelH{sec:setup}Setup}
A few examples of our aggregates, made of 1000 monomers each, are
presented in \fg{sample}. These particles \cd{consist of} equal size
monomers.
\figlarge{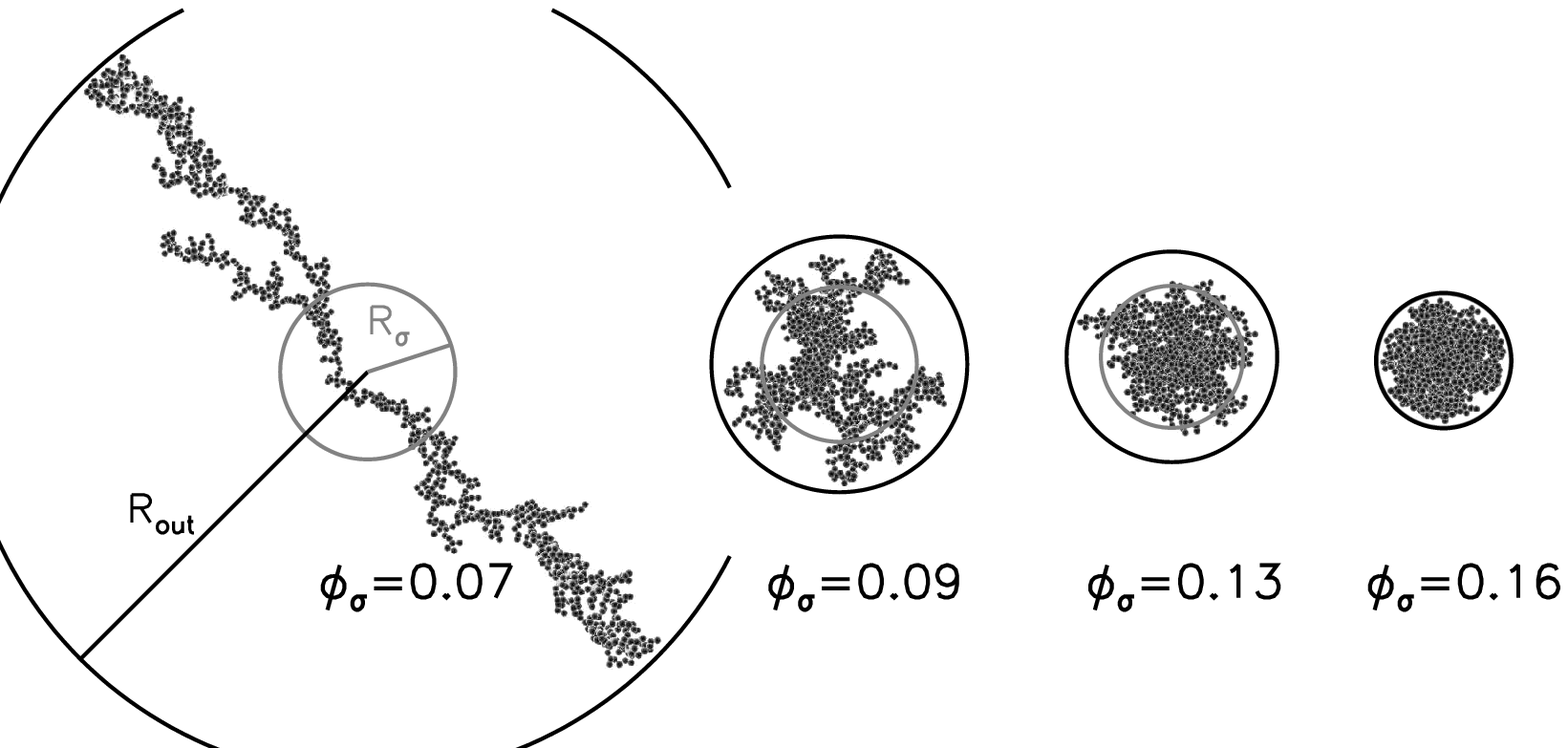}{Sample aggregates used in our
  parameter study.  Each of particles is made of 1000 monomers.
  Annotations show \tpsig for each particle.  \cd{Black circles}
  define the outer radius $R_\mr{out}$, while gray circles
  indicate the projected surface equivalent radius $R_\sigma$ (see
  text).}{sample}
\cd{These aggregates} are constructed using two different techniques.
The first method, developed by
\citet{2000JournCollInterfSci...229..261F}, allows to create fractal
aggregates of any specified fractal dimension $D_\mr{f}$. This
\emph{sequential tunable particle-cluster aggregation method} forms
agglomerates by successive addition of identical spherical
particles.  The fractal aggregates made according to this algorithm
\cd{strictly} obey the fractal scaling law
\citep{2000JournCollInterfSci...229..261F}
\begin{equation}
N=k_\mathrm{f}\Biggl(\frac{R_\mathrm{g}}{a}\Biggr)^{D_\mathrm{f}},
\labelH{eq:fractal-scaling-law}
\end{equation}
where $R_\mathrm{g}$ is the radius of gyration and $k_\mr{f}$ is the
fractal pre-factor that is related to the central packing density. For
the aggregates used in our study, we choose a pre-factor
$k_\mr{f}=1.6$ and fractal dimensions of $D_\mr{f}=1.5$,
$D_\mr{f}=2.0$, and $D_\mr{f}=2.5$, \cd{respectively.}  \cd{We measure
  the filling factor of aggregates by considering a sphere with radius
  $R_{\sigma}$, having a cross section $\pi R_{\sigma}^2$ which is
  equal to the projected surface of the aggregate, averaged over all
  angles (see section~\ref{compactness-par}).  We then compute the
  filling factor $\psig$ as the ratio of the solid volume in the
  aggregate and the volume of the equivalent sphere.  With this
  definition, aggregates of the mentioned fractal dimensions, made of
  1000 monomers, have filling factors $\psig=0.07$, $\psig=0.09$, and
  $\psig=0.16$, respectively.}

The second method we use to produce aggregates is \cd{particle cluster
  aggregation} (PCA).  We successively add monomers from random
directions.  This \cd{method} produces aggregates that, in the limit
of very large sizes, have a filling factor of $\psig=0.15$.  An
aggregate made of 1000 monomers has, however, a lower filling factor
of $\psig=0.13$. This is an effect of high porosity of the surface
layer. In very large aggregates, this region will be negligibly thin
compared to the size of an aggregate
\citep{2008PaszunHierarchical_inprep}.

\subsection{\labelH{sec:paramspace}Parameter space \emph{and expected scaling}}
To provide a qualitative and quantitative description of aggregate
collisions, we explore an extensive parameter space.  This provides an
insight into the effects of different parameters on the outcome of a
collision.  As the final purpose is to provide a recipe for a collision
between two aggregates, we limit the range of our parameters to
realistic values.

\cd{The relevant parameters influencing the outcome of a collision are:}
\begin{itemize}
\item{collision energy,}
\item{pre-collision compactness of the aggregate,}
\item{mass ratio of colliding particles,}
\item{impact parameter,}
\item{material properties of monomers.}
\end{itemize}
The values used for the different parameters are shown in
\tb{parameter-space}.  The columns in the table refer to (1)-relative
collision velocity $v$, (2)-pre-collision filling factor of an
aggregate \tpsig, (3)-impact parameter 5 $b/b_\mr{max}$, and (4)-mass
ratio of colliding aggregates $m_1/m_2$.  Although not all
combinations are simulated, each parameter set used it computed for
six different orientations in order to obtain a fair average of
possible collision outputs.
\begin{table}
\caption{Impact parameters and aggregate properties explored in this
  study.}
\labelH{tab:parameter-space}
\begin{tabular}{c c c c c }
\hline\hline
$v$ [m/s] & \tpsig & $b/b_\mr{max}$ & $m_1/m_2$ \\
(1) & (2) & (3) & (4) \\
\hline
$0.01$   & 0.07  & 0.0   & $10^{-3}$  \\ 
$0.05$   & 0.09  & 0.25  & $5\ep{-2}$ \\ 
$0.1$    & 0.122 & 0.5   & $1\ep{-1}$ \\ 
$0.2$    & 0.13  & 0.75  & $2\ep{-1}$ \\ 
$0.3$    & 0.155 & 0.875 & $4\ep{-1}$ \\ 
$0.5$    & 0.16  & 0.95  & $6\ep{-1}$ \\ 
$0.75$   & 0.189 &       & $8\ep{-1}$ \\
$1.0$    & 0.251 &       & $1$        \\ 
$2.0$    &       &       &            \\
$4.0$    &       &       &            \\ 
$6.0$    &       &       &            \\ 
$8.0$    &       &       &            \\ 
$10.0$   &       &       &            \\
\end{tabular}
\end{table}

\subsubsection{Collision energy}
The \cd{basic} effect of impact energy is intuitive: The more energy
is provided to the system, the more violent the outcome
is. \citet{1997ApJ...480..647D} and later \citet{2007Apj...661..320W}
provide a simple recipe of collisional output as a function of energy.
Both these studies were limited to two dimensional aggregates.  Their
recipe predicts energy thresholds for processes such as erosion,
compression, and fragmentation.  To understand them, one needs to
define the rolling energy and the breaking energy of a contact. The
first one is the energy needed to roll two monomers over each other by
90 degrees, and it represents energy the related to restructuring of
an aggregate. It is defined as \citep{1997ApJ...480..647D}
\begin{equation}
E_\mr{roll}=6\pi^2\gamma R \xi_\mr{crit},
\labelH{eq:rolling-energy}
\end{equation}
where $\gamma$ is a surface energy, $R$ is reduced radius of the two
monomers in consideration, and $\xi_\mathrm{crit}$ is a critical
displacement at which the rolling becomes irreversible and energy is
dissipated.  From measurements of the rolling friction between spheres
it was shown that this quantity should be set to about $20$\AA\ in
order to reproduce the measured energy \cd{losses
  \citep{1999PhRvL..83.3328H}.}

The second important quantity is the energy needed to separate two
connected monomers and is defined as
\begin{equation}
E_\mathrm{br}=1.8 F_\mathrm{c} \delta_\mathrm{c}.
\labelH{eq:critical-energy}
\end{equation}
Here $F_\mathrm{c}$ is the pull-off force, minimum force needed to
disconnect two monomers and it is
\begin{equation}
F_\mathrm{c}=3 \pi \gamma R
\labelH{eq:pull-off-force}
\end{equation}
where $\delta_\mathrm{c}$ is the critical displacement from the equilibrium
position \citep{1997ApJ...480..647D}.  Thus, two monomers in contact,
when pulled apart, will separate at this position. $\delta_\mr{c}$ is
defined as
\begin{equation}
\delta_\mathrm{c}=\frac{1}{2} \frac{a_0^2}{6^{1/3}R},
\labelH{eq:critical-displacement}
\end{equation}
with the equilibrium contact radius
\begin{equation}
a_0= \biggl( \frac{9 \pi \gamma R^2}{E^*} \biggr)^{1/3}.
\labelH{eq:a0}
\end{equation}
Putting these equations together we see that the critical energy is given by
\begin{equation}
E_\mr{br} = A \frac{\gamma^{5/3} R^{4/3}}{E^{\ast 2/3}},
\labelH{eq:generic-critical-energy}
\end{equation}
with the dimensionless constant $A=43$.

The recipe of \citet{1997ApJ...480..647D} is confirmed experimentally with one
important modification \citep{2000Icar..143..138B}. The energy scaling should be
applied according to values determined empirically (i.e. $\xi_\mr{crit}\approx
20\mr{\AA}$ and $E_\mr{br}$ corresponding to the experimental results).  In
order to satisfy this requirement, we introduce a scaled version of
$E_\mr{br}$.  In this case the constant $A$ is higher such that the sticking
threshold in our model is in agreement with experiments
\citep{2000ApJ...533..454P}. This energy is given by
\begin{equation}
E_\mr{br} = 2.8\ep{3} \frac{\gamma^{5/3} R^{4/3}}{E^{\ast 2/3}}.
\labelH{eq:breaking-energy}
\end{equation}

The recipe by \citet{1997ApJ...480..647D} is summarized in \tb{recipe-DT}.
\begin{table}
\caption{The collision recipe from \citet{1997ApJ...480..647D} for a 2D case.}
\labelH{tab:recipe-DT}
\begin{tabular}{r|l}
\hline \hline 
Energy & Outcome of Collision \\ 
\hline 
$E_\mr{impact} < 5 E_\mr{roll}$ & Sticking without restructuring \\ 
$E_\mr{impact} \approx 5 E_\mr{roll}$ & Onset of restructuring local \\
& to the impact area \\ 
$E_\mr{impact} \approx n_\mr{c} E_\mr{roll}$ & Maximum compression \\ 
$E_\mr{impact} \approx 3 n_\mr{c} E_\mr{br}$ &Onset of erosion (start to \\
& lose monomers)\\ 
$E_\mr{impact} > 10 n_\mr{c} E_\mr{br}$ & Catastrophic disruption \\ 
\hline
\end{tabular}
\end{table}
Low energies are insufficient to cause any visible restructuring.
Before any restructuring occurs, contacts between monomers are
displaced elastically.  \cd{Only when the critical displacement is
  exceeded, an irreversible motion (i.e. rolling or sliding) follows.}
When the energy \cd{is sufficient to roll several contacts by a}
significant angle ($\sim$90 degrees), monomers in the impact region
begin to roll and visible restructuring occurs.  The amount of
restructuring increases with increasing energy.  Maximum compression
is typically reached, when all monomers have enough energy to roll by
90 degrees.

Monomers begin \cd{to be removed of an aggregate} when the energy per
contact reaches 3 times the breaking energy. A large fraction of the
energy is then dissipated by rolling, but the excess is used to remove
a few grains. As the impact energy increases, the amount of erosion
increases and becomes catastrophic when each contact can access over
10 times the breaking energy.

Although this recipe is reasonable, it may differ when particles
collide at an impact parameter other than $b=0$ or when the mass ratio
of the two collision partners is not 1.  The latter case was also
considered by \citet{1997ApJ...480..647D}.  They provided a recipe for
collisions of a small grain with a cluster.

In this study we also explore the energy range from the hit-and-stick
regime up to catastrophic destruction.  \cd{This range of energies can
  be found by looking at a variety of astrophysical environments.  For
  example, in protoplanetary disks, small particles are well coupled
  to the gas and have rather low relative velocities.  In the cores of
  molecular clouds, however, low gas densities make it possible for
  particles to decouple from the gas even at small sizes.  In this
  case, turbulent gas motions may lead to collisions velocities beyond
  the fragmentation limit \citep{2008OrmelMolcloud_inprep}}.

\subsubsection{Mass ratio}
The mass ratio is sampled within the available range of our model
($10^{-3} < m_1/m_2 < 1$).  As the largest aggregate simulated in this
study is made of 1000 monomers, the lowest mass ratio \cd{we consider}
is $m_1/m_2=10^{-3}$ (a collision of a big aggregate with a monomer).
Although we simulate collisions between particles with different mass
ratios, we present only two limiting cases to \cd{illustrate} the
importance of the mass ratio.

This parameter influences the energy distribution during a
collision. In the case of equal-mass impactors, the energy is spread
over most of the monomers, causing a global effect.  However, small
projectiles act \emph{locally} and concentrate the energy into a small
volume close to the impact site.  \cd{Thus, restructuring will be local
and erosion can be expected at relatively low energies.}

\subsubsection{Impact parameter}
Our sampling of the impact parameter $b\ $ covers the range from a
central impact up to a \emph{grazing collision}, where the impact
parameter $b$\ equals 95\% of the sum of the outer radii of two
colliding aggregates $R_\mr{out,1}+R_\mr{out,2}$.  We do not consider
impact parameters above 95\% as the irregular shape of aggregates
generally causes such ``collisions'' to miss.

The impact parameter is important as it can significantly change the
outcome of a collision. In the case of \cd{a} central impact,
aggregates are pushed towards each other and are compressed.
Collisions with large impact parameters on the other hand cause
aggregates to connect in the outer regions only and result in
\emph{stretching} of aggregates as they move apart again.  \cd{In this
  way,} large impact parameters tend to cause tensile forces acting on
the aggregates, while central impacts are dominated by compressive
stresses.

The importance of the \cd{off-center} collisions should not be
underestimated.  Due to the \cd{geometrical} arrangement, impacts with
large impact parameter are considerably more frequent than \cd{head-on
  collisions}. \cd{In impact} parameter averaging, grazing impacts
have a lot of weight.

\subsubsection{Compactness parameter}
\label{compactness-par}
To describe the structural changes of aggregates we define a compactness
parameter \tpsig\ as
\begin{equation}
\psig=N \biggl( \frac{r_0}{R_\sigma} \biggr)^3, \labelH{eq:phisigma}
\end{equation}
where $N$ is the number of monomers in the aggregate, $r_0$ is a
monomer radius, and $R_\sigma$ is the projected surface equivalent
radius (cf. \fg{sample}), defined as
\begin{equation}
R_\sigma = \sqrt{\frac{\sigma}{\pi}},
\labelH{eq:Rsigma}
\end{equation}
with $\sigma$ being the projected surface averaged over many
orientations.  The inverse of this filling factor was introduced
earlier by \citet{2007A&A...461..215O} as the \emph{enlargement
  factor} $\psi$.  Note that this parameter alone is insufficient to
fully describe the structure of an aggregate.  An additional quantity
is required to avoid ambiguity, e.g. the fractal dimension.  However,
for small aggregates such as the ones in this study, the possibilities to
distribute mass within the particle are limited.

We sample the compactness parameter well within the applicable range.
Random close packing (RCP) of \cd{spheres produces aggregates} with
the filling factor of about\footnote{Although the value of 0.635
  correspond to filling factor defined in respect to the outer radius
  $R_\mr{out}$ and not $R_\sigma$, both radii are equal for these
  compact aggregates.} $\psig \approx 0.635$
\citep{1990PhRvL..64.2727O} that is the densest form one can expect,
given the assumption of spherical monomers.  However,
\citet{2006ApJ...652.1768B} and later \citet{2008A&A...484..859P}
\cd{have shown} that aggregates being compressed can reach a maximum
filling factor of about $\psig = 0.33$. Higher compaction cannot be
achieved in a static experiment of uni-axial compression.  As upper
limit for the filling factor we use a slightly lower value of
$\psig=0.25$. As lower limit, on the other hand, we use aggregates
formed in the Brownian growth phase, where in the presence of rotation
aggregates with the fractal dimension of about $D_\mr{f}=1.5$ are
formed \citep{2004PhRvL..93b1103K,2006Icar..182..274P}.  Our largest
aggregate of fractal dimension 1.5 has a filling factor $\psig=0.07$.

It is important to note the discrepancy between the radii used in the
definition of the impact parameter space and the filling factor
\tpsig. The impact parameter is defined in terms of the outer radius
$R_\mr{out}$, that is the radius of a sphere enclosing an entire
aggregate and centered in its center of mass. The filling factor,
however, uses the projected surface-equivalent radius $R_\sigma$.  For
compact aggregates, these two radii are very similar, while the outer
radius $R_\mr{out}$ becomes higher as the filling factor of an
aggregate decreases.  For completeness, we empirically determine the
relation between the two radii. \Fg{Rout-Rsig} shows how the ratio
$R_\mr{out}/R_\sigma$ influences the filling factor \tpsig.
\figsmall{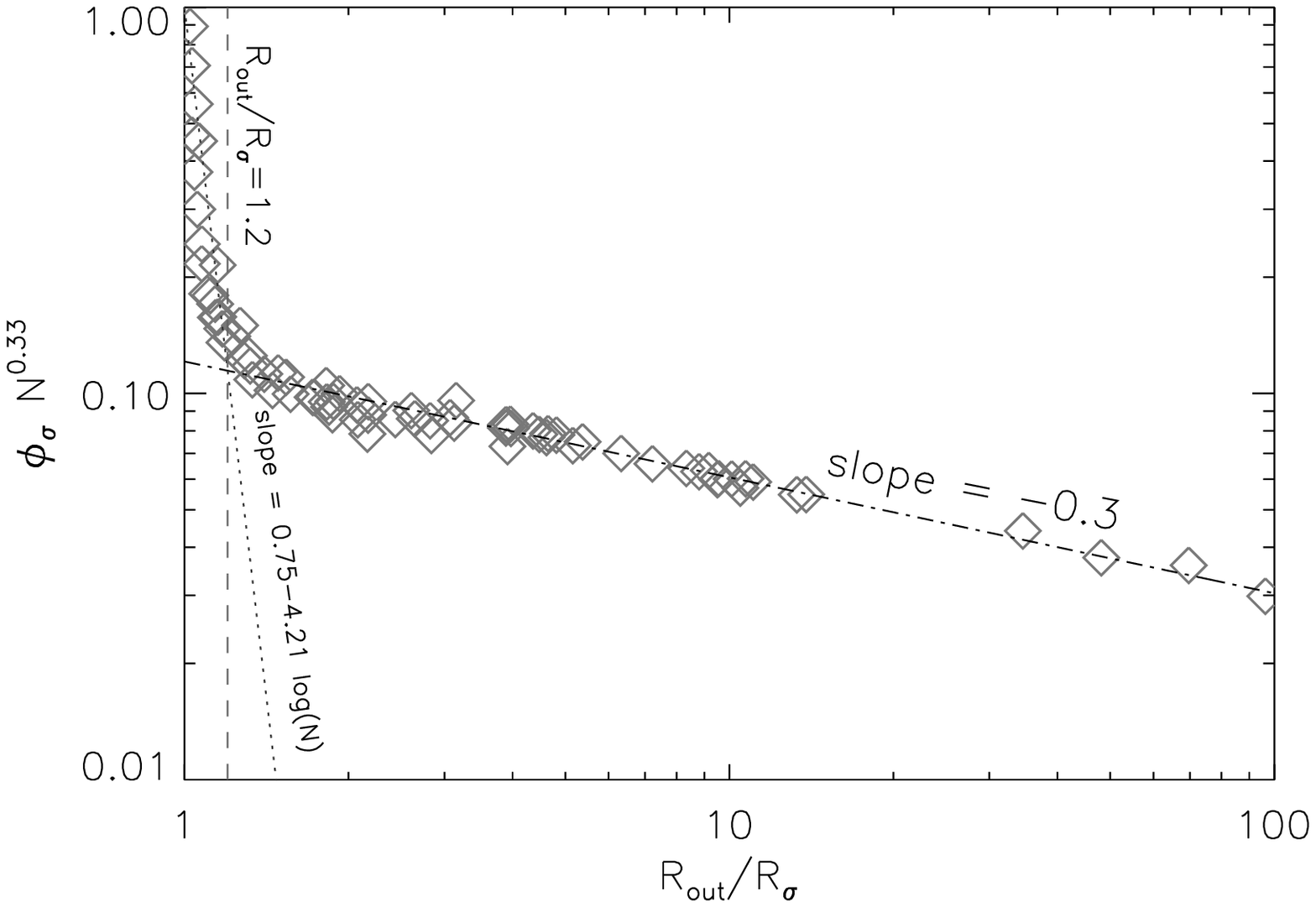}{The filling factor as a function
  of the ratio $R_\mr{out}/R_\sigma$ of the outer radius of an
  aggregate over its projected surface equivalent radius. Each point
  corresponds to an aggregate (see text) of different structure and
  mass and thus different ratio of the two radii. The dashed-dotted
  line shows the least square fit of the power-law to this data. A
  sharp transition between compact and fluffy particles occurs at
  about $R_\mr{out}/R_\sigma=1.2$ (dashed line). Compact aggregates
  have a smaller ratio of the two radii.  Their filling factor is
  approximated by a \cd{separate} power-law (dotted line).}{Rout-Rsig}
To obtain this relation we constructed (using
\eq{fractal-scaling-law}) many aggregates of different structures
($D_\mr{f}$, $k_\mr{f}$) and masses ($N$). Interestingly all data is
very well confined along a simple curve. As the filling factor seems
to depend on the mass of aggregate as $\psig \propto N^{-0.33}$, we
plot $\psig N^{0.33}$ in \fg{Rout-Rsig} to collapse all data on a
single curve. This mass dependence is further discussed in
\se{results-ff}. Compact particles in \fg{Rout-Rsig}
($R_\mr{out}/R_\sigma<1.2$) have a similar outer \cd{radii}
$R_\mr{out}$ and projected surface equivalent \cd{radii} $R_\sigma$.
They show little dependence of the $R_\mr{out}/R_\sigma$ ratio on the
filling factor.  Fluffy aggregates ($R_\mr{out}/R_\sigma>1.2$) on the
other hand show a power-law relation of the filling factor on the
ratio of the two radii with a slope of $-0.3$.  The fitted power-law
is presented in \fg{Rout-Rsig} as dashed-dotted line.  The complete
relation that holds for ratio $R_\mr{out}/R_\sigma > 1.2$ is given as
\begin{equation}
\psig = 1.21 \left( \frac{R_\mr{out}}{R_\sigma} \right)^{-0.3} N^{-0.33}.
\labelH{eq:Rout-Rsig}
\end{equation}

\subsubsection{Properties of monomers}
The composition of monomers and their size strongly affect the
strength of an aggregate.  Both these parameters determine the
breaking energy of two grains in contact and thus regulate the energy
dissipation during a collision.  Since normalization of the impact
energy by the breaking energy puts monomer properties out of the
equation, we study here only one monomer size and material. The
physical parameters of Quartz used in our model are presented in
\tb{material-properties}. Other materials will be the subject of a
future study.
\begin{table}[!h]
\caption{Properties of monomers used in this study. Material properties
  correspond to Quartz monomers or radius $r_0=6\ep{-5}$ cm.}
\begin{tabular}{c c c c}
\hline\hline
$\gamma$ [erg/cm$^2$] & $E^\ast$ [dyn/cm$^2$] & $\xi_\mr{crit}$ [cm]  & $\rho$ [g/cm$^3$]  \\ 
\hline
$25$ & $2.78\ep{11} $ & $2\ep{-7}$ & 2.65 \\
\end{tabular}
\labelH{tab:material-properties}
\end{table}
Our monomers are silica spheres with a diameter of 1.2 \um.  The
elasticity modulus $E^\ast$ is defined as
\begin{equation}
E^\ast = \biggl( \frac{1-\nu_1^2}{E_1} + \frac{1-\nu_2^2}{E_2} \biggr)^{-1},
\labelH{eq:elasticity}
\end{equation}
where $E_i$ and $\nu_i$ are the Young's modulus and the Poisson ratio
of $i$-th monomer, respectively.

%
%
%
%
\section{\labelH{sec:results}Results and discussion}
In this section we present results of our parameter study. We describe
the collisional outcome in terms of the fragment \cd{mass}
distribution - growth versus fragmentation.  Moreover, we keep track
of the structure of the fragments formed in such a collision and
present the collisional evolution of the structure of dust aggregates.

Our study \cd{spans a wide range} of parameters.  Here we select two
specific cases for a detailed discussion of the effects seen in our
study\footnote{The quantitative recipe presented in \se{recipe} does
  of course make use of the entire parameter study.}.  Collisions of
compact aggregates (filling factor of $\psig=0.251$ for smaller
aggregates and $\psig=0.16$ for larger particles) are presented and
compared with fluffy, fractal aggregates (filling factor of
$\psig=0.155$ for small aggregates and $\psig=0.09$ for bigger
ones). We use these two cases to illustrate the influence of
compactness on the collisional outcome.

\subsection{\labelH{sec:results-fd}Fragment distribution}

The products of a collision are generally quantified in terms of the
fragment mass distribution.  In the case of sticking, the resulting
mass distribution contains one single element, with the mass given by
the sum of impactor and projectile.  With the onset of erosion, a
second component appears - the distribution of small fragments,
usually represented as a power-law of particle mass.  While these are
initially two clearly separated components, they can connect in
increasingly destructive collisions.

\subsubsection{The effect of impact energy}

\paragraph{Basic mass distribution components}

Although the impact energy seems to be the main quantity setting the
mass spectrum after the collision, the collisional outcome depends
very much on all parameters presented in \se{paramspace}. The largest
collisional remnant is presented as a function of the impact energy in
\fg{remnant}.  At low collision energies, aggregates stick perfectly
and the largest fragment contains the mass of both colliding
particles.  An increasing impact energy causes onset of erosion at the
energy of about $E=0.1NE_\mr{br}$. The fragmentation occurs at the
energy of about $E\sim NE_\mr{br}$, and depends also on the internal
structure of the colliding aggregates and the impact parameter.

Compact aggregates with densely packed monomers ($\psig=0.251$) can
sustain higher energies regardless of the impact parameter. Grains
packed close to each other undergo more interaction, resulting in
internal energy dissipation that is more efficient than in the
case of loosely packed monomers ($\psig=0.155$).
\figsmall{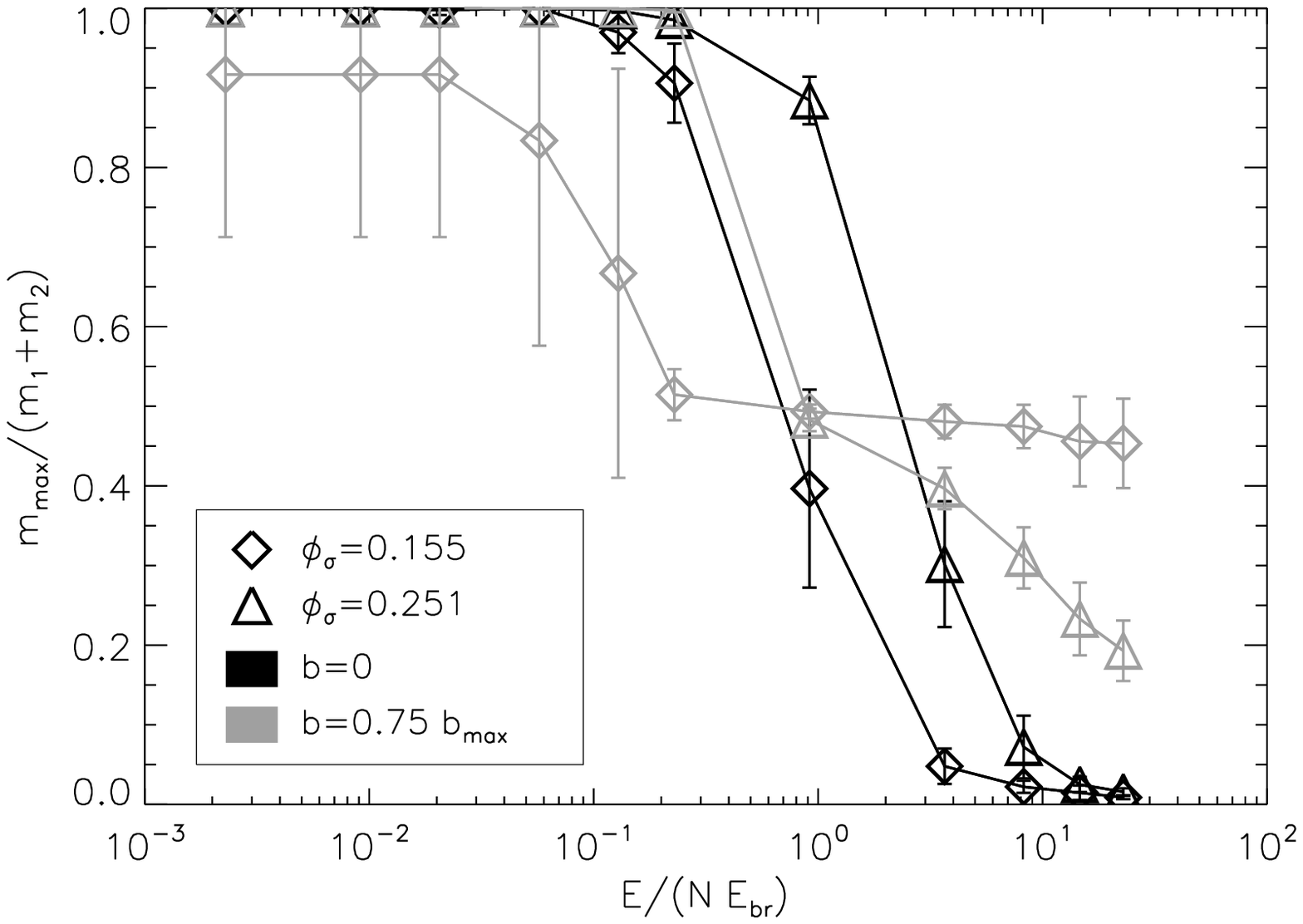}{The mass of the largest collision
  remnant for central and offset collisions as a function of the
  impact energy. Diamonds correspond to compact aggregates, while
  triangles correspond to fluffy particles.  Error bars indicate
  standard deviation and each point represents a mean value from 6
  individual simulations (see \se{setup}).}{remnant}

Interestingly, erosion occurs in off-center collisions already much
earlier than in central collisions, while shattering has the opposite
behavior: It occurs more readily in central collisions.  Central
impacts can cause catastrophic destruction at energies above
$E=10NE_\mr{br}$, while off-center collisions at this same energy
usually cause two large fragments to remain and therefore must be
classified as be classified as erosion.  Both effects are caused by
\cd{the} distribution of the available impact energy within \cd{the}
colliding aggregates.  In off-center collisions, the energy is
concentrated \cd{into a} small region of the aggregates and does break
grain-grain connections \cd{locally near the} point of impact.  The
remaining kinetic energy is then carried away by two massive
fragments.  In the limiting case of $(b=b_\mr{max})$ the interaction
of the two aggregates occurs between only two monomers, if at all.

The degree of fragmentation during off-center impacts depends then on
the packing density of grains in the two aggregates.  Porous
aggregates ($\psig=0.155$) show that even at very high energies, the
energy is not efficiently absorbed by the aggregates, saving them from
being shattered.  However, in the case of compact aggregates, the mass
of the largest collision remnant is significantly higher than in the
case of central impact, but it decreases with an increasing impact
energy.

\paragraph{The full mass distribution}

A more complete picture is presented by studying the full
distribution of fragment masses. This is illustrated for central
impacts in \fg{distro-inter}. Collisions at
\figsmall{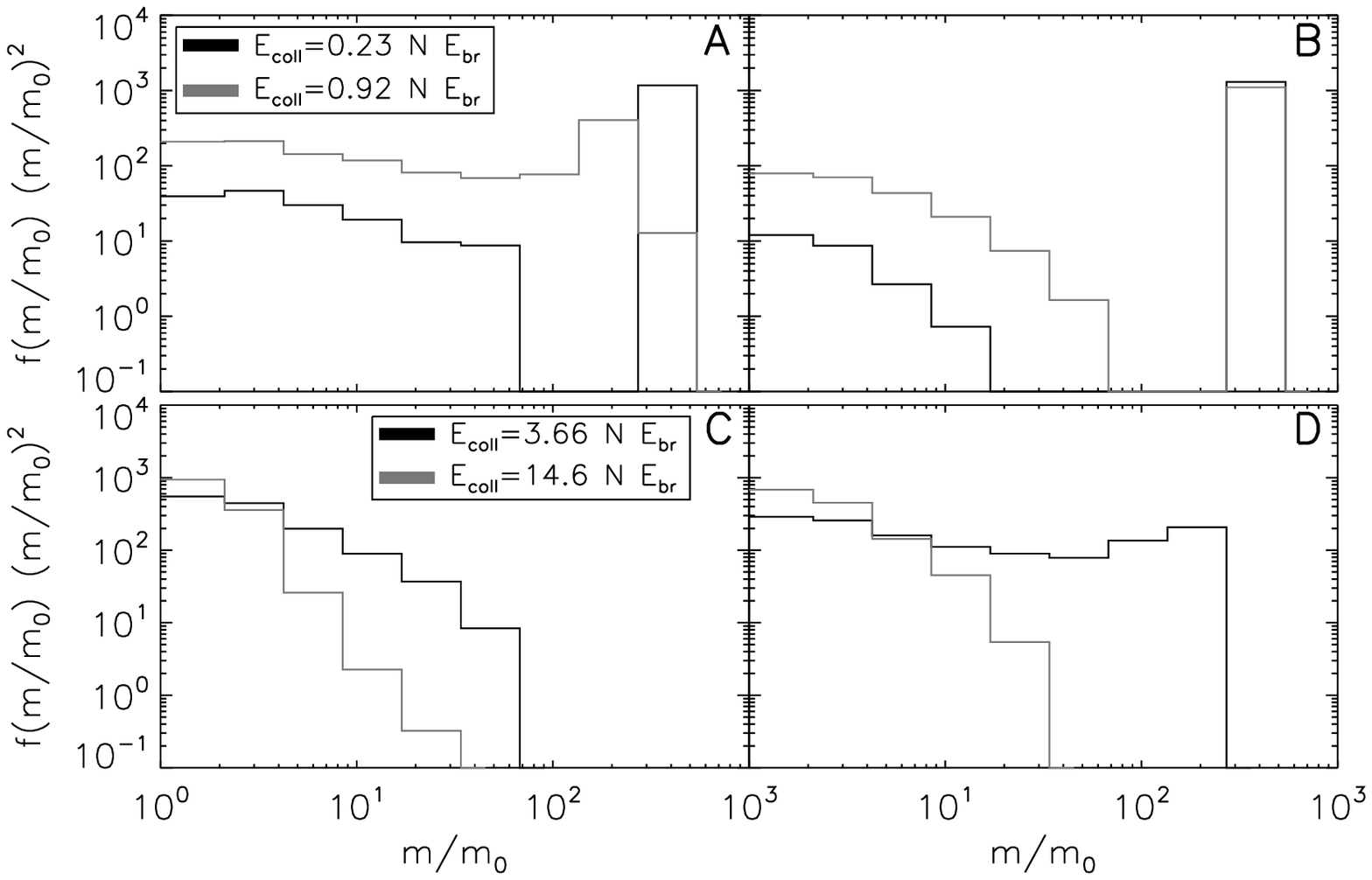}{Mass
  \cd{distribution resulting} from head-on collisions of A\&C --
  fluffy ($\psig=0.155$) and B\&D -- compact ($\psig=0.251$)
  aggregates at four different impact energies.}{distro-inter}
intermediate speeds (top panels) erode aggregates and produce a number
\cd{of small} fragments.  Both the power-law and the largest fragment
components are present.  \cd{Depending on} the degree of erosion they
can be separated or connected.  However, the slope of the power-law
remains more or less unchanged.

The distribution of small particles is defined as
\begin{equation}
f(m/m_0)(m/m_0)^2 \propto (m/m_0)^q,
\labelH{eq:mpwl-distro}
\end{equation}
with the slope $q$ depending on several parameters, including the
internal structure of aggregates and the impact energy.  Initially,
too few small particles are produced and the power-law cannot be
determined (\eg for off-set impacts see \fg{distro-inter-b3}a\&b). In
such a case we assume that the distribution is flat with the slope
$q=0$, meaning equal mass per logarithmic mass interval.  As the
erosion progresses with increasing impact energy, the fitted slope
remains at an approximately constant low value.  Small particles
slightly dominate the mass within this power-law distribution, since
the slope is about $q \approx -0.3$ in the case of fluffy aggregates
(\fg{distro-inter}a), and about $q \approx -1.2$ for compact
aggregates (\fg{distro-inter}b).  These slopes remain constant up to
energies of about $E=NE_\mr{br}$.  Faster impacts provide
\cd{sufficient energy} to break all contacts in an
aggregate.  Therefore, full fragmentation sets in, which can
significantly alter the fragment distribution.  The head-on collision
at energy of $E=3.66 N E_\mr{br}$ completely shatters fluffy
aggregates with the largest fragment having only about 15\% of the
total mass.  The large fragment component disappears altogether,
shifting the entire mass into the power-law component with the now
steeper slope $q\approx -1$.  Further increase in the energy results in
heavier damage and a still steeper slope.  At the energy of $E=14.6 N
E_\mr{br}$ the slope reaches $q\approx-2.5$ (see \fg{distro-inter}c).
The steepening slope can be interpreted as monomers and very small
fragments in the fragment distribution becoming dominant.

Although a similar trend is observed for compact particles
($\psig=0.251$), the fragmentation is not as effective as in the case
of aggregates with $\psig=0.155$ (see \fg{distro-inter}d). A head-on
impact of compact particles at the energy of $E=3.66 N E_\mr{br}$
\cd{can still be classified as erosion}.  The large fragment component
is still present and contains a significant fraction of the mass.
This component is broader and is connected with the power-law
distribution of small fragments, \cd{an effect which affects} the
determination of the slope.  \cd{The slope appears to decrease} to
about $q \approx -0.3$. However, an increase in the collision energy
to $E=14.6 N E_\mr{br}$ shatters the aggregates, leaving only the
power-law component with a steeper slope of about $q\approx -1.7$ (see
\fg{distro-inter}d).

The relation of the energy and the slope of the power-law distribution
in central collisions is presented in \fg{slopes}. The left panel (a)
shows the results for aggregates \cd{with compactness} parameter
$\psig=0.155$.  \cd{The initially shallow and constant slope begins}
to steepen once the impact energy increases above $E= N E_\mr{br}$.
This behavior is similar for aggregates of different masses \cd{or for
  different mass ratio impacts.}  Note that for very weak erosion, the
slope of the power-law cannot be determined and a value of $q=0$ is
assumed. \cd{Shattering} becomes catastrophic when the \cd{impact
  energy is increased} by an order of magnitude.  The slope steepens
beyond $q=-1$, meaning that small \cd{particles dominate} the mass
spectrum.  For small aggregates the distribution may be as steep as
$q=-2.5$.
\figsmall{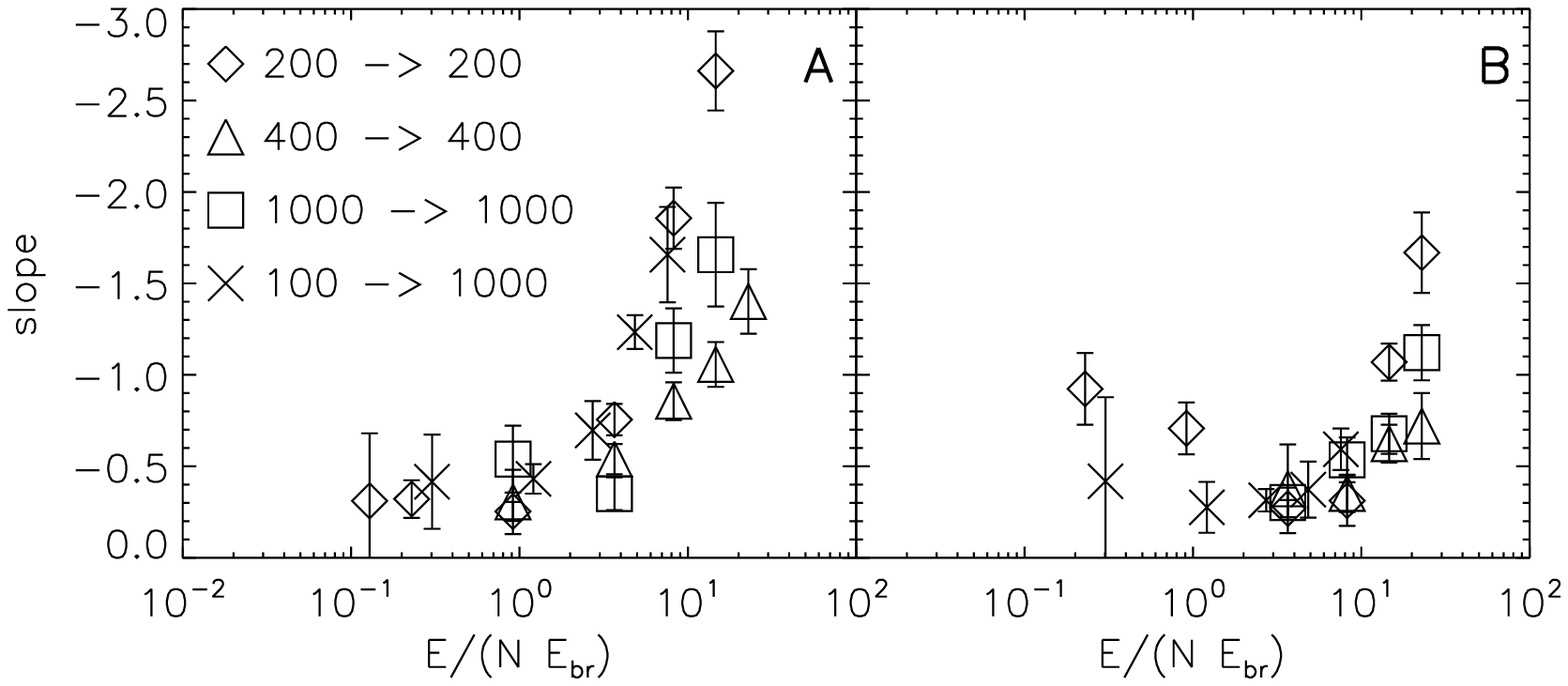}{\cd{Slopes of the power-law size
    distribution of small fragments as a function of impact
    energy. Different symbols correspond to different masses of the
    colliding aggregates.  The left panel} (A) corresponds to fluffy
  aggregates ($\psig=0.09 \ldots 0.155$), while the right panel (B)
  shows \cd{the} result of collisions between compact aggregates
  ($\psig=0.16 \ldots 0.251$).}{slopes}

Compact particles behave in a similar manner (see \fg{slopes}b). In
this case, however, the steepening occurs at slightly higher energies
of about $E=5 N E_\mr{br}$.  Moreover, the steepening is limited in
the explored energy range, and reaches values of $q=-1.7$.  In this
case, low energies also result in shallow slopes of the distribution
and are \cd{again} assumed to be $q=0$ when the data was too scarce to
make a fit.

\subsubsection{The effect of impactor-to-target mass ratio}

Below we discuss the effect of the mass ratio on the collision
outcome.  Although intermediate energies are required to erode
particles in collisions of equal mass aggregates, this is not the case
for impacts with high mass ratios,
%
\figsmall{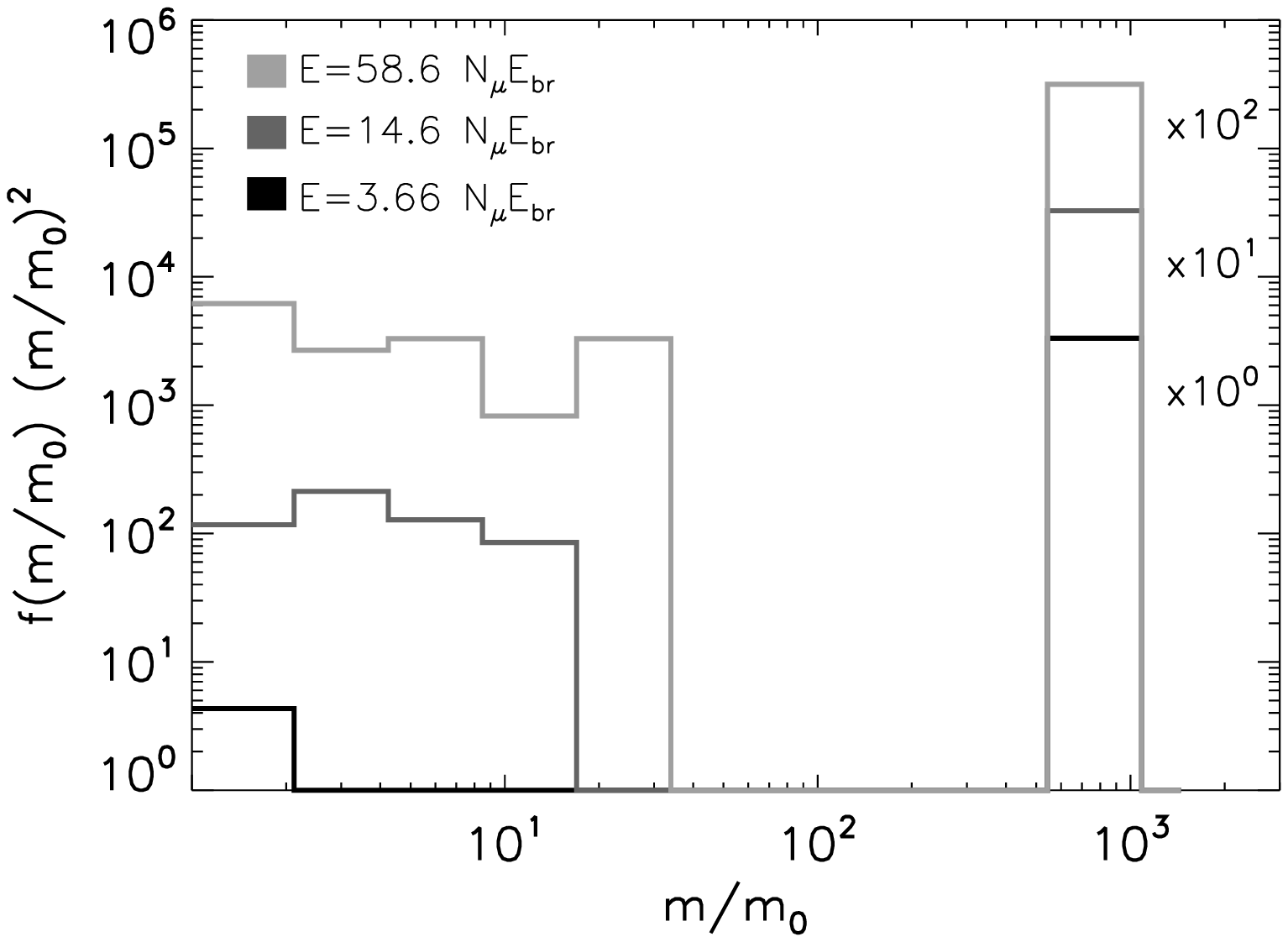}{Mass
  distribution \cd{resulting} from central collisions between a
  monomer and an aggregate made of 1000 grains. The presented
  distributions result from collisions with compact aggregate with the
  filling factor of $\psig=0.16$.  Note that the energy is normalized
  to the reduced number of monomers
  $N_\mu=N_1N_2/(N_1+N_2)$.}{distro-mass-ratio-panel}
where the impact energy is localized to a small region, leading to
erosion even in slow collisions. \Fg{distro-mass-ratio-panel} shows
the fragment distribution produced in collision of a monomer with an
aggregate made of $10^3$ grains. In these cases, the energy is
sufficient to break only a low number of contacts\footnote{Note that
  in \fg{distro-mass-ratio-panel} the energy is normalized to the
  reduced number of monomers $N_\mu=\frac{N_1N_2}{N_1+N_2}$.
  Therefore, the energy per contact is a factor of $10^3$ lower than
  in the case of a collision of particles of equal mass.} and results
in erosion.  That same energy applied in a collision of equal mass
aggregates results in a perfect sticking without any mass loss
(cf. \fg{remnant}).  This difference is a consequence of very
localized energy input.  The small particle (in this case a monomer)
carries sufficient energy to break a number of contacts.  This energy
is \cd{transmitted} locally to a limited number of grains rather than
distributed over \cd{the} entire target aggregate.  \cd{The resulting
  ejecta can, due to the small physical size of the projectile, easily
  escape.}

Similarly, off-center collisions (see \fg{distro-inter-b3}) are
characterized by distributions that resemble the erosion case.  A
highly pronounced large fragment component coexists with a power-law
distribution of small fragments.  The slope of the power-law is
independent of the impact energy.  It remains at about $q\approx-1$
for fluffy aggregates and $q\approx-0.6$ for compact particles even at
very high energies above $E=10 N E_\mr{br}$.  At low impact energies,
the resulting distribution are
\figsmall{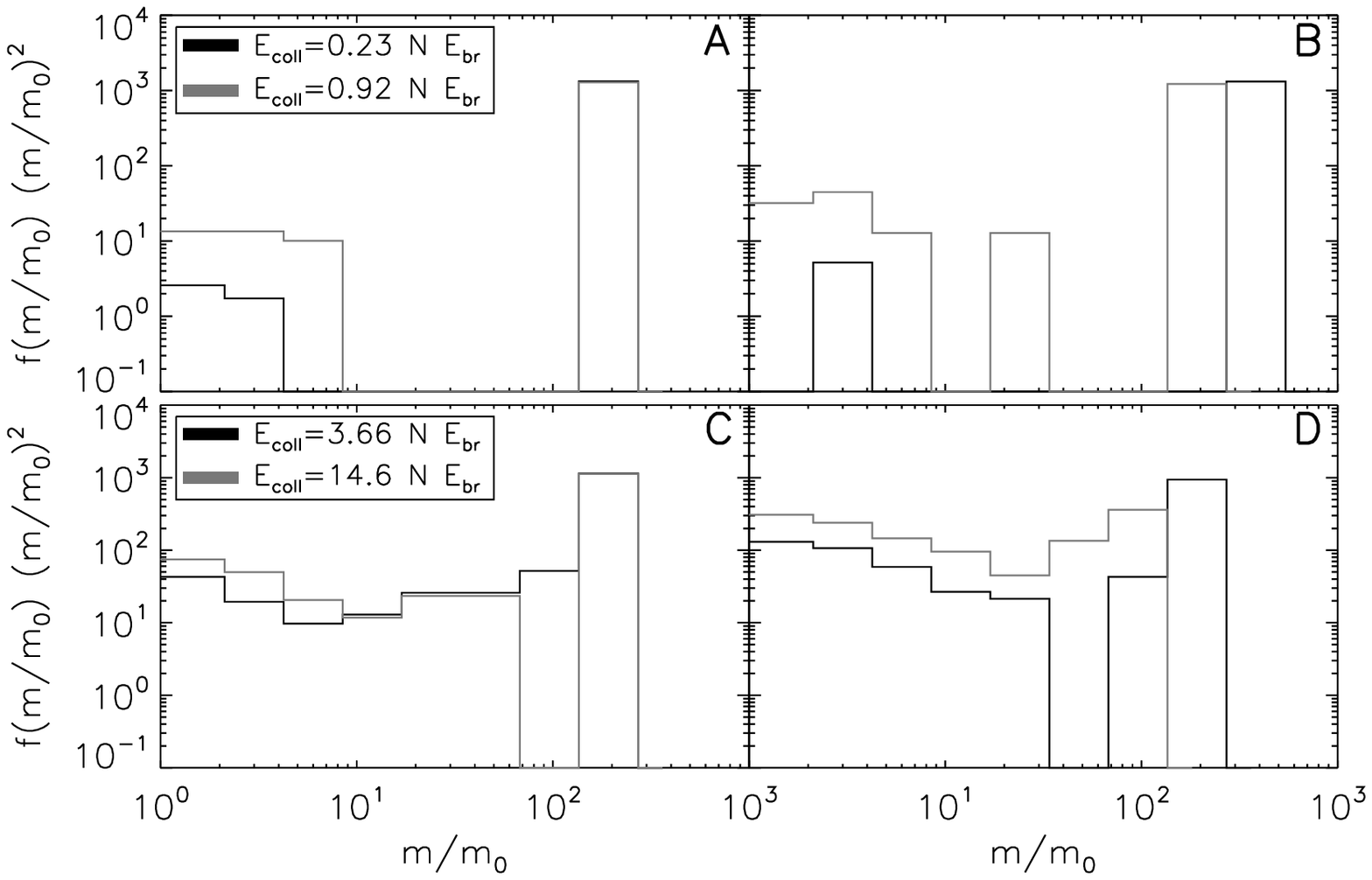}{Mass distribution as
  resulted from off-center ($b=0.75\ b_\mr{max}$) collisions of A\&C -- fluffy
  ($\psig=0.155$) and B\&D -- compact ($\psig=0.251$) aggregates at four
  different impact energies.}{distro-inter-b3}
dominated by the largest fragment component and the slope of the
power-law component \cd{is undetermined due} to the low number of
fragments.

Fluffy aggregates, when colliding \cd{at an energy of $E = 0.23 N
  E_\mr{br}$ and above}, produce two large fragments that contribute
to the big particles component and a few small particles.  Compact
aggregates can still grow at an energy of $E= 0.23 N E_\mr{br}$,
producing one large fragment containing most of the mass.  At energies
of about $E\approx N E_\mr{br}$ the erosion increases and the growth
is stopped.  The large fragment component contains now two large
remnants of the initial aggregates.  \cd{An increase} of the collision
energy causes mainly an increase in the mass in the power-law
component, while the largest fragment remains weakly affected. In the
case of compact aggregates ($\psig=0.251$), only very high energy
impacts can significantly reduce the mass of the largest fragments
(see \fg{remnant} and \fg{distro-inter-b3}d). For aggregates with
lower packing density ($\psig=0.155$), the mass of the largest
collision remnant seems to stabilize at a \cd{value} of about
$m_\mr{max}\approx 0.5 (m_1+m_2)$ once the impact energy exceeds $E=N
E_\mr{br}$.  Very high energies influence the fragment distribution
very weakly (see \fg{remnant} and \fg{distro-inter-b3}c).

The main effect of the impact parameter is that the energy is not
transported very efficiently into aggregates. In the case of a central
collision the kinetic energy is naturally transported efficiently into
both aggregates, as the interaction spreads from the center (\ie
region right in between the aggregates) outwards.  The monomers are
pushed closer together, actively taking part in the energy
dissipation.  An increase in the impact parameter results in a
decrease in the number of \cd{actively interacting grains}.  This
means that fewer grains actually collide resulting in less
fragmentation.

Our results show that the main factor \cd{determining} the collisional
outcome is the impact energy $E$.  Other parameters influence the way
this energy is transported to and distributed over the available
monomers. \Fg{sketch-fragment-distro} shows \cd{a} schematic picture
of how the fragment distribution changes with variation of impact
energy $E$, compactness parameter $\phi_\sigma$, impact
parameter $b$, and the mass ratio $m_1/m_2$.  The arrows indicate
schematically how the position and scaling of the different components
of the fragment mass distributions shift as parameters vary.
\figsmall{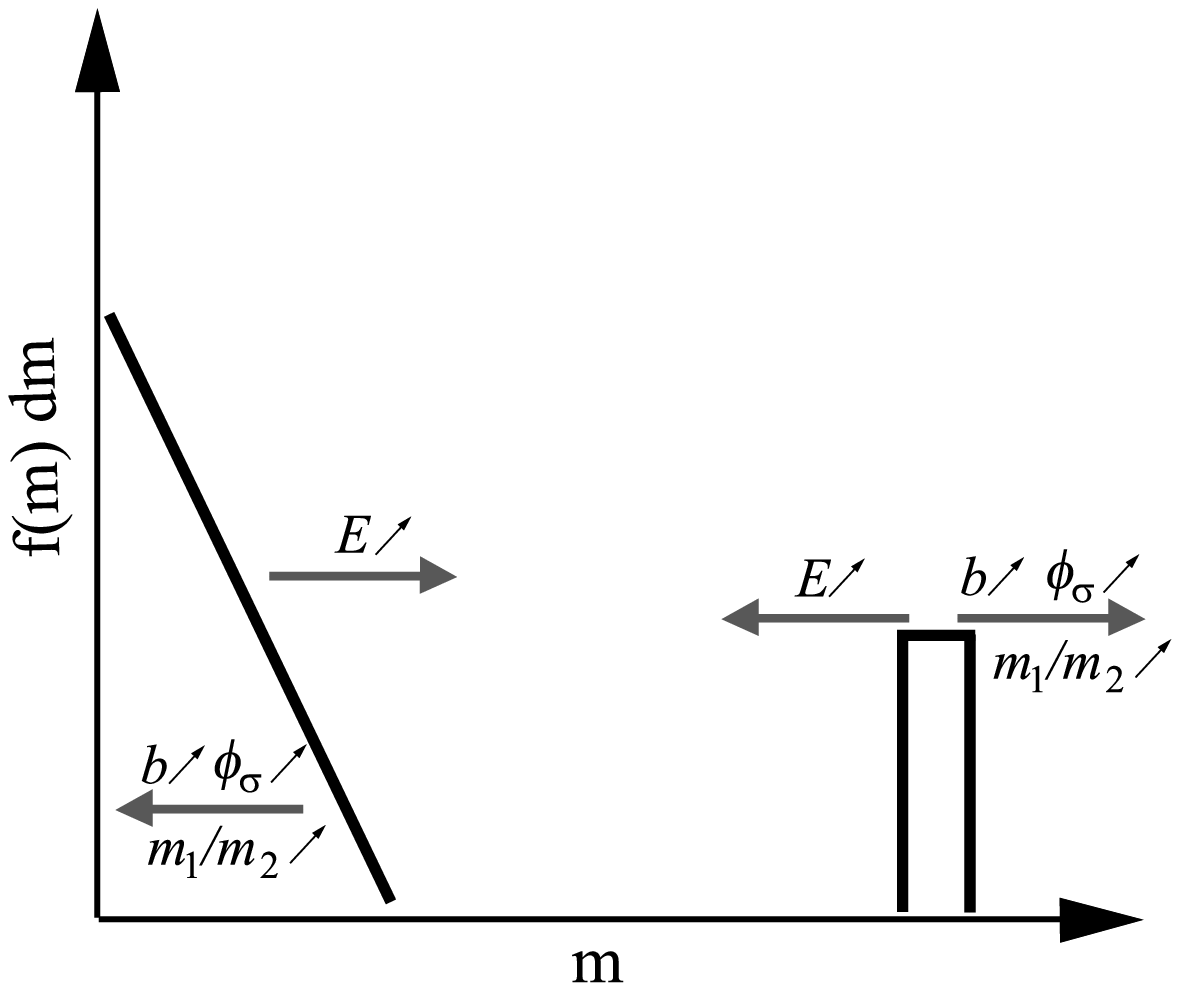}{Schematic plot of the
evolution of the fragment distribution as a function of different
parameters.}{sketch-fragment-distro}
%



\subsection{\labelH{sec:results-ff}Aggregate structure}

\subsubsection{Central collisions}
The outcome of a collision can also be quantified in terms of the
internal structure of the resulting aggregates.  This quantity is very
important as it determines the \cd{aerodynamic} properties of
aggregates and thus their relative velocities. \Fg{panel20} shows
examples of the effect of collisions involving aggregates with
$\psig=0.155$.
\figlarge{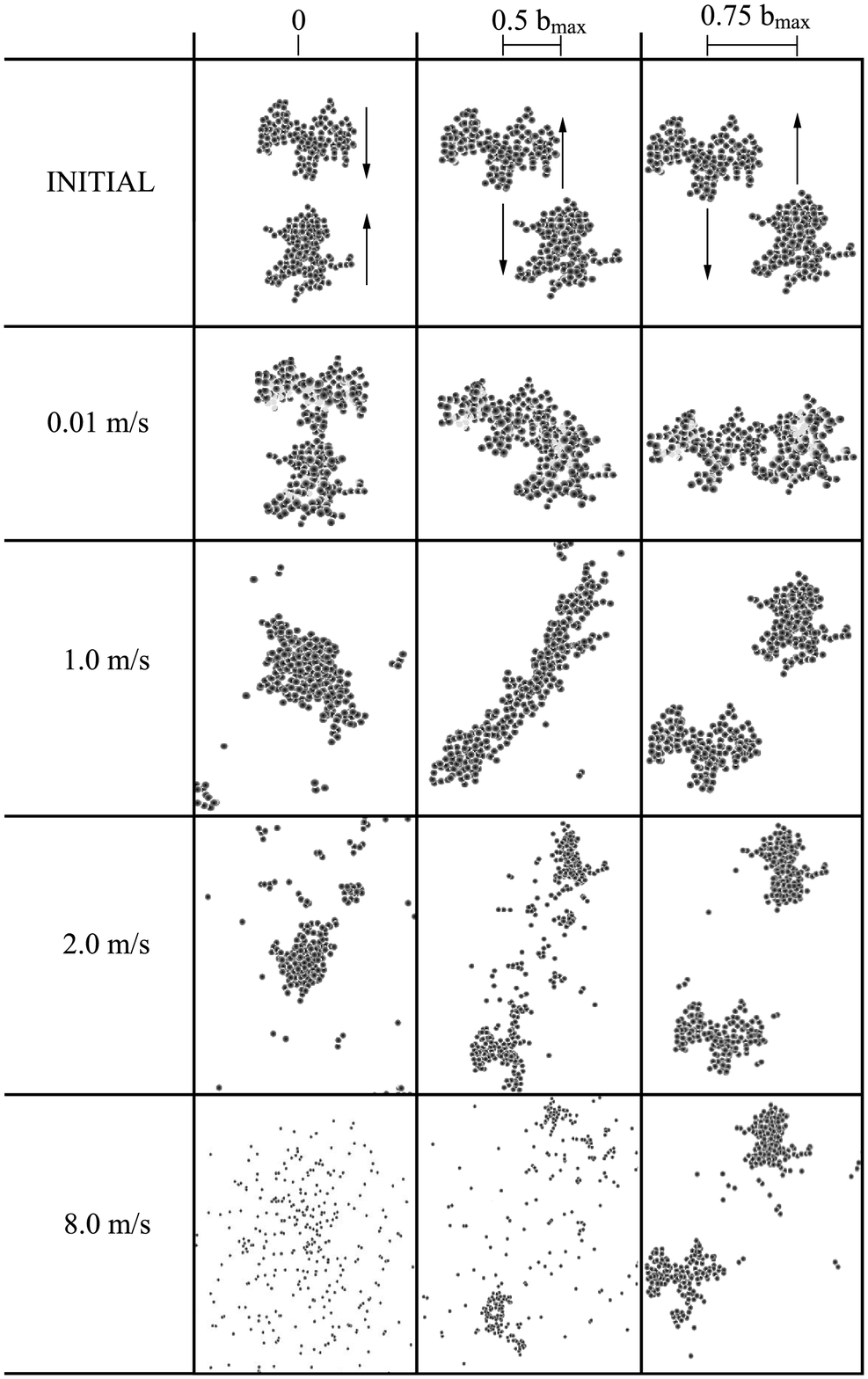}{Images of aggregates formed in
  collisions at different impact energies $E$ and at different impact
  parameters $b$. \cd{The aggregates} have an initial filling factor
  of $\psig = 0.155$.  Top panels show the initial setup.  Lower
  panels present result of a collision. Images are in different scales
  to show different processes, i.e. compaction or
  fragmentation.}{panel20}
Central collisions evolve from the hit and stick regime through
compaction and flattening to fragmentation.  As the impact parameter
increases, this path changes.  Then the hit and stick regime is
followed by the \emph{stretching} regime, where aggregates are pulled
into an elongated shape, exposing more surface in the process.
Eventually, sufficient energy is reached to break fragile connection
between the colliding aggregates, at the location of the highest
stress.  At this moment, two aggregates of similar mass are created.
Depending on the impact parameter and energy, erosion may also occur
in the vicinity of the interaction area.

This picture depends also on the initial compactness of \cd{the}
colliding aggregates.  Particles with open structures (\fg{panel20})
are subject to drastic restructuring, while more compact ones are
characterized by higher strength against compacting and tensile
forces. For comparison we present the results
\figlarge{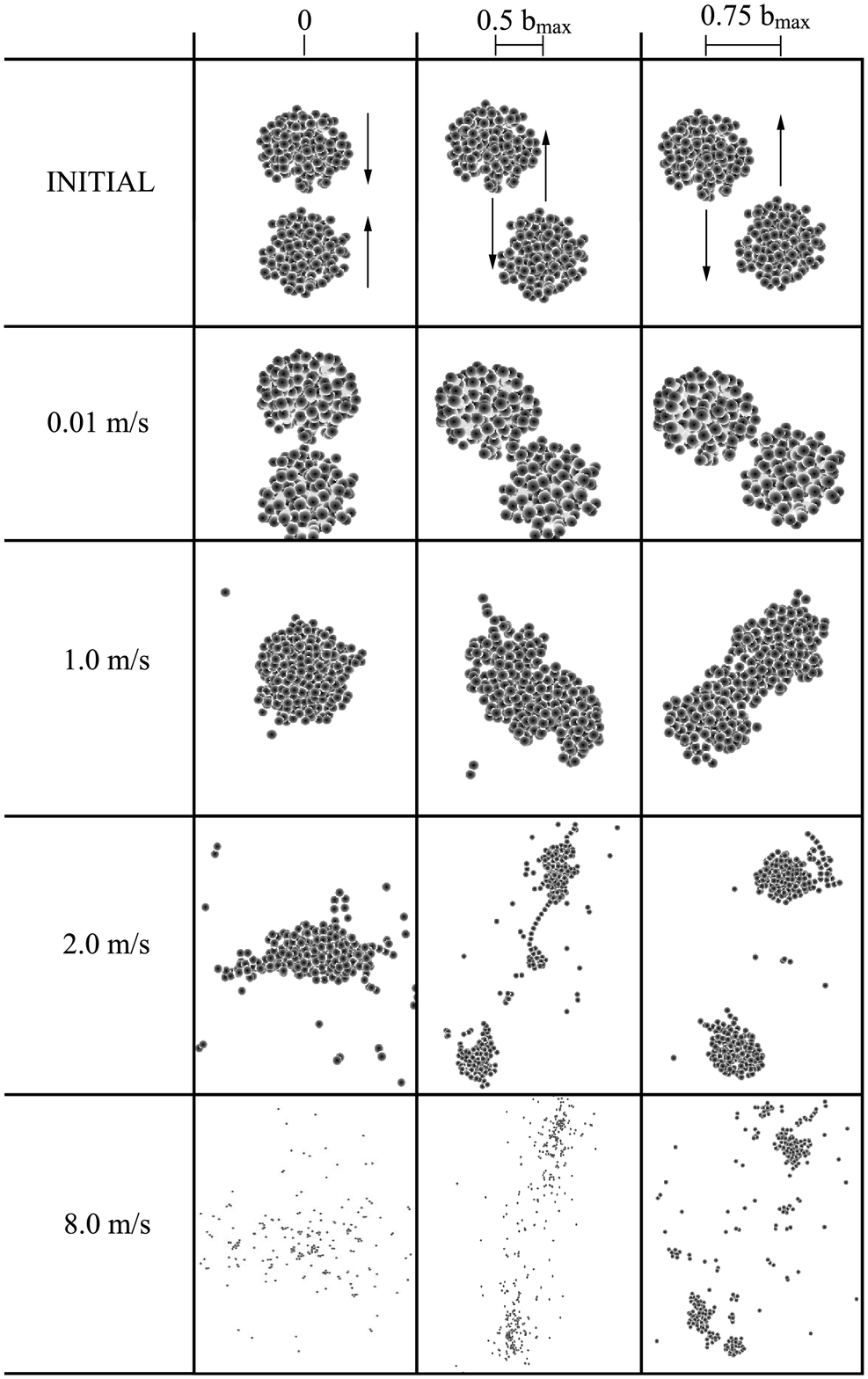}{Images of aggregates formed in collisions at
different impact energies $E$ and at different impact parameters $b$. Aggregates
have the filling factor of $\psig = 0.251$.  Top panels show the initial
setup. Lower panels present result of a collision. Images are in different scales
to show different processes, i.e. compaction or fragmentation.}{panel25}
for aggregates with geometrical filling factor of $\psig=0.251$ in
\fg{panel25}.  In this case the produced aggregate is not stretched as
much as in the case of more fluffy aggregate.  Additionally, the erosion
is much stronger at higher impact parameters. This is caused by the
closer packing of monomers that enables more efficient energy transfer
into the aggregates.

The quantitative picture is presented in \fg{av-final}.  The
geometrical filling factor is plotted as a function of the impact
energy for central and off-center
\figsmall{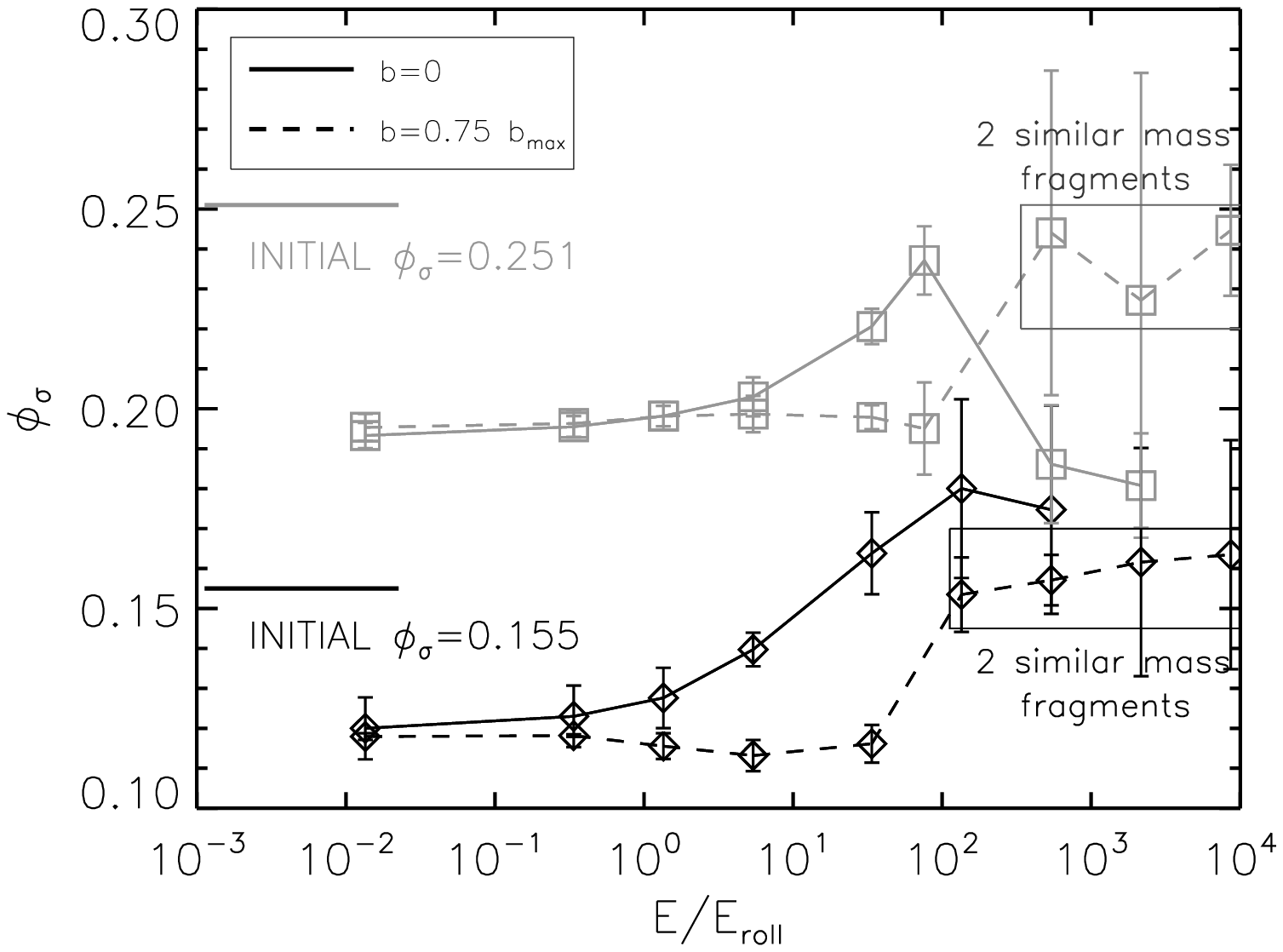}{The geometrical filling factor $\psig$
as a function of impact energy for head-on (solid lines) and off-center (dashed
lines) collisions. Black lines correspond to aggregates with lower initial
filling factor of $\psig = 0.155$, while gray lines correspond to higher
filling factor of $\psig = 0.251$.}{av-final}
%
%
collisions.  All post-impact aggregates are in the hit and stick
energy regime, \cd{below} a few times the rolling energy
$E_\mr{roll}$. This is consistent with \cd{the} description provided
by \citet{1997ApJ...480..647D}, where the visible restructuring occurs
at \cd{impact energies} above $5\,E_\mr{roll}$. However, the structure
of the produced aggregate is different than that of \cd{the} initial
particles.  Aggregates stick, forming \emph{dimer-like} structure.
Thus, they expose more surface area resulting in a decrease in the
geometrical filling factor.  This decrease depends on the initial
structure of aggregates and \cd{appears} to be more severe for
\cd{compacter} aggregates.  In this case, particles cannot penetrate
each other contrary to more open structure aggregates, where partial
overlap hides some surface area. In fact, the difference is rather
small, on the order of 20\%, for both filling factors presented in
\fg{av-final}.

Further increase in the impact energy, beyond $5\,E_\mr{roll}$,
results in an \emph{increase} of the filling factor.  In the case of
central collision aggregates undergo compression up to energy of about
$0.5\, N\, E_\mr{roll}$.  At this energy, the maximum compaction is
reached and this energy is close to threshold predicted by
\citet{1997ApJ...480..647D} and confirmed experimentally by
\citet{2000Icar..143..138B}.  The degree of compression depends on the
initial compactness, as compact aggregates are more difficult to
compress further. An aggregate with $\psig=0.155$ reaches its maximum
compression of about $\psig=0.18 \pm 0.02$, more compact than the
pre-impact particles.  A more compact aggregate with pre-impact
filling factor $\psig=0.251$ does not exceed it's initial compactness
after impact.  The maximum compression obtained in our study is
somewhat lower than obtained in a quasi-static compression
\citep{2006ApJ...652.1768B, 2008A&A...484..859P}. This, however, can
be an effect of small aggregate sizes, where the filling factor is
strongly affected by the porous outer layers
\citep{2008PaszunHierarchical_inprep}.

A further increase in the collision energy causes flattening and,
formally, a \emph{decompaction} of aggregates.  Even thought the
filled regions of the aggregates continue to show low porosity, the
non-spherical global shape exposes more surface than a spherical
structure would, leading to a decrease in the geometrical filling
factor.  In the case of fluffy aggregates with $\psig=0.155$, the
flattening results in a small decrease in the compactness.  At \cd{a}
collision energy of about $N\, E_\mr{roll}$ the maximum decompaction
is reached and any further increase in the impact energy would lead to
fragmentation.  For more compact aggregates with initial $\psig=0.251$,
however, the flattening is stronger, as the filling factor drops to
about $\psig \approx 0.18$.  This filling factor can is similar to
what we have seen earlier in fluffy aggregates.  Fragmentation is now
only be reached at about $3\,N\,E_\mr{roll}$.

\subsubsection{The influence of the impact parameter}

Off-center collisions, on the other hand, show different results.
Energies that lead to compaction in central impacts also cause
restructuring at large impact parameters. In this case, however,
particles are pulled apart and stretched.  Thus, more surface area is
exposed resulting in a strong decrease in the geometrical filling
factor \tpsig. The stretching energy regime extends, however, to lower
energies \cd{than the compression regime}.  Aggregates connect with
\cd{a} lower number of contacts that are pulled off. Therefore,
lower energy is sufficient to disconnect the two colliding aggregates.
The critical energy \cd{is about} $0.25\, E_\mr{roll}$, and may be
slightly higher for compact aggregates.  Above that energy, two
particles of similar mass are produced, accompanied by erosion.  All
small fragments produced both in central and off-center collisions
closely follow a single power-law relation of
\begin{equation}
\psig = (m/m_0)^{1/3}.
\label{eq:psig-small}
\end{equation}
The reason for this consistent behavior lies in the fact that the
fragments are produced in highly dynamical events with energies close to
breakup energies, allowing internal restructuring to lead to a
scale-free structure.

\subsubsection{Schematic representation}

\Fg{sketch-av} sketches a general picture of the structural evolution
of aggregates.  On the two axes, in arbitrary units, it shows filling
factor and mass.  We assume that initially, both impactor and target
have equal properties, located at the center of the plot where the
solid black curve starts.  Along that curve, the impact energy
increases in steps, and the position of points on that curve shows
possible structures of the largest post-impact aggregate.  The first
segment shows the effect of a hit-and-stick collision, to double mass
and lower filling factor, independently of impact parameter.  At that
point, the curve splits into two, for central (solid black) and
grazing (gray) \cd{collisions}.  Higher impact energies can then, with
equal mass, either increase or decrease the filling factor.  Even
higher energies move to the third point on each path.  \cd{Grazing
  collisions return to the initial pre-impact properties,} while
central collisions go through a maximum compression point onto the
fragmentation powerlaw $\psig = (m/m_0)^0.33$.

\figsmall{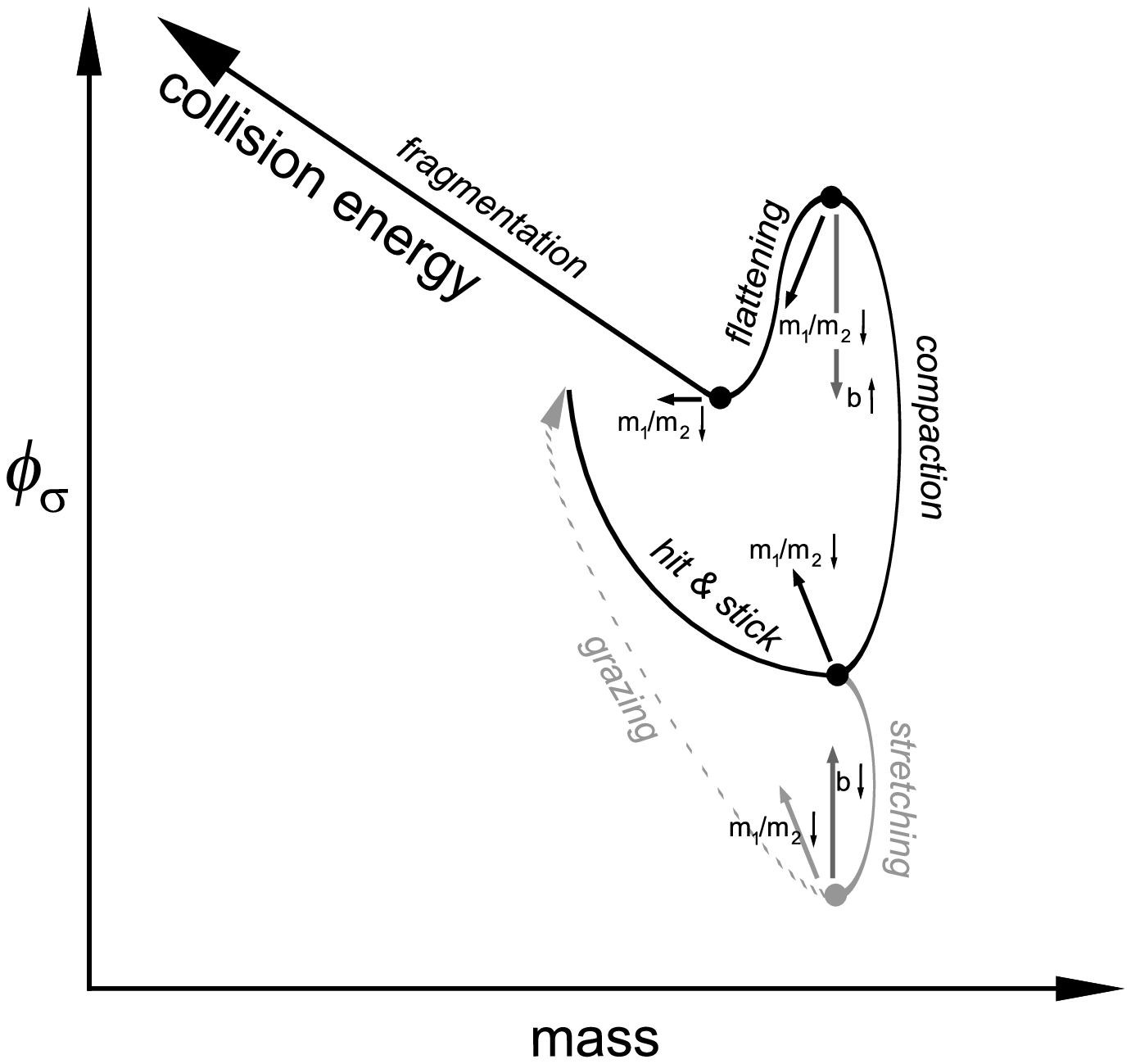}{Sketch of the compactness evolution
dependence on different parameters.  See text.}{sketch-av}
Small arrows in the sketch indicate how the different points shift
around as model parameters are modified.

%
%
%
%
\section{\labelH{sec:recipe}The Recipe}
In this section we put together the information regarding the
collisional output in a form of a quantitative recipe.  We provide the
recipe in a form of tables that contain parameters needed to
reconstruct appropriate distributions.  Since the recipe is provided
as discrete parameters, the intermediate cases should be
interpolated. Our recipe describes two limiting cases separately.
Equal mass collisions affect aggregates globally.  Therefore we refer
to this sub-recipe as the global recipe.  On the other hand we have
high mass ratio impacts that result in localized changes.  This
sub-recipe is referred to as the local recipe.  In the case of large
mass ratio collisions, we used impacts of monomers onto aggregates
composed of 1000 grains.  In this case, only central collisions were
used \cd{to predict also the outcome of offset} impacts, because the
effects are extremely similar.  This assumes that the geometry of the
impact of small projectile onto a larger target depends on a local
surface. Thus central and offset collisions should be
indistinguishable.  The only necessary correction is to exclude missed
collisions that are more likely at larger impact parameters.  The
distinction between the two sub-recipes is described later in
\se{global-local}.

\subsection{\labelH{sec:b-average}Average over the impact parameter}

\cd{We present our recipe for aggregate collisions in an
  impact-parameter-averaged way.  In this way, the recipe can be
  easily used in models that deal with size and structure
  distributions rather than individual particles.  It is also possible
  to rewrite the recipe, \cd{keeping the impact parameter as} one of
  the collisional parameters.  This can be useful in a Monte-Carlo
  approach \citep{2008OrmelMolcloud_inprep} where individual
  collisions are treated.}

Since we know the products of collisions at different offsets, we
apply proper weights to \cd{the} results.  The weight for each impact
parameter $b$ is related to the fractional surface area of the ring
with the radius $b$ and width $\Delta b$.  Thus, grazing collisions get
the highest weight, while the central impact has the lowest
weight. The quantity $Q$ averaged over the impact parameter is then
given as
\begin{equation}
<Q>_b = \frac{\int_0^{b_\mr{max}} Q(b) 2 \pi b\ \mr{d}b}{\pi b_\mr{max}^2}.
\labelH{eq:b-average}
\end{equation}

\subsection{Hit and stick recipe}
Although our simulations include the hit-and-stick energy regime, this
growth mechanism strongly depends on the mass ratio and should be
treated separately.  The hit-and-stick growth is of little importance
when only an insignificant mass is added before any restructuring
occurs. Otherwise, an analytical prescription may be applied as in
\citet{1993A&A...280..617O,2007A&A...461..215O,2008OrmelMolcloud_inprep,2008PaszunHierarchical_inprep}.

\subsection{Distribution of fragment masses}

The simulation results have shown that collisions produce two
components.  The first component - the power-law of small fragments -
is produced in high energy, head-on impacts resulting in erosion and
fragmentation.  The second component - the distribution of the largest
fragments - is formed during low energy impacts, and also by grazing
collisions at all energies.  We model this component using (somewhat
arbitrarily) a \cd{Gaussian distribution of particle masses}.  The
overall effects seen in the previous chapter indicate that, as the
energy increases, the Gaussian component will move its peak position
to smaller masses, the powerlaw component will gain relative
importance and will eventually steepen.

The total set of parameters needed to reproduce such a distribution is
presented in \tb{recipe-parameters}. The power-law component is
determined by fitting a
\begin{table}
\caption{Quantities provided by the recipe to reproduce the mass distribution.}
\begin{tabular}{r l}
\hline \hline
 symbol         & description                                  \\
\hline
$q$             & The slope of the power-law component.        \\
$M_\mr{r}$      & ratio of the mass in the power-law component \\
                & to the mass in the Gaussian component.       \\
$\sigma_\mr{G}$ & Width of the Gaussian component.             \\
$M_\mr{G}$      & Mean mass of the Gaussian component.         \\
\hline
\end{tabular}
\labelH{tab:recipe-parameters}
\end{table}
power-law to the first part of the distribution, which contains small
fragments.  That power-law is then subtracted from the distribution.
The power-law slope and mass ratio of the two components is known at
this point. The remaining part of the distribution is \cd{then} used
to determine the last \cd{two} quantities, namely the mean and the
width of the Gaussian component.  The mean mass $M_\mr{G}$ is simply
the mass-weighted mean mass of fragments in the Gaussian component
\begin{equation}
M_\mr{G} = \frac{\int_{\log m_0}^{\log (m_1+m_2)}m f(m/m_0) (m/m_0)^2\mr{d}\log
  m}{\int_{\log m_0}^{\log (m_1+m_2)} f(m/m_0) (m/m_0)^2 \mr{d}\log m}.
\labelH{eq:mean-gaussian}
\end{equation}
 The width of the Gaussian is chosen to obey two constraints:
\begin{enumerate}
\item{the Gaussian should have a sharp cut-off at masses larger than the total
  mass of colliding aggregates.}
\item{the power-law component must dominate the low mass part of the
  distribution.}
\end{enumerate}

To obtain a representative fragment distribution, we average the mass
spectrum over the impact parameters (see \se{b-average}).  Some
impact-parameter-averaged
\figlarge{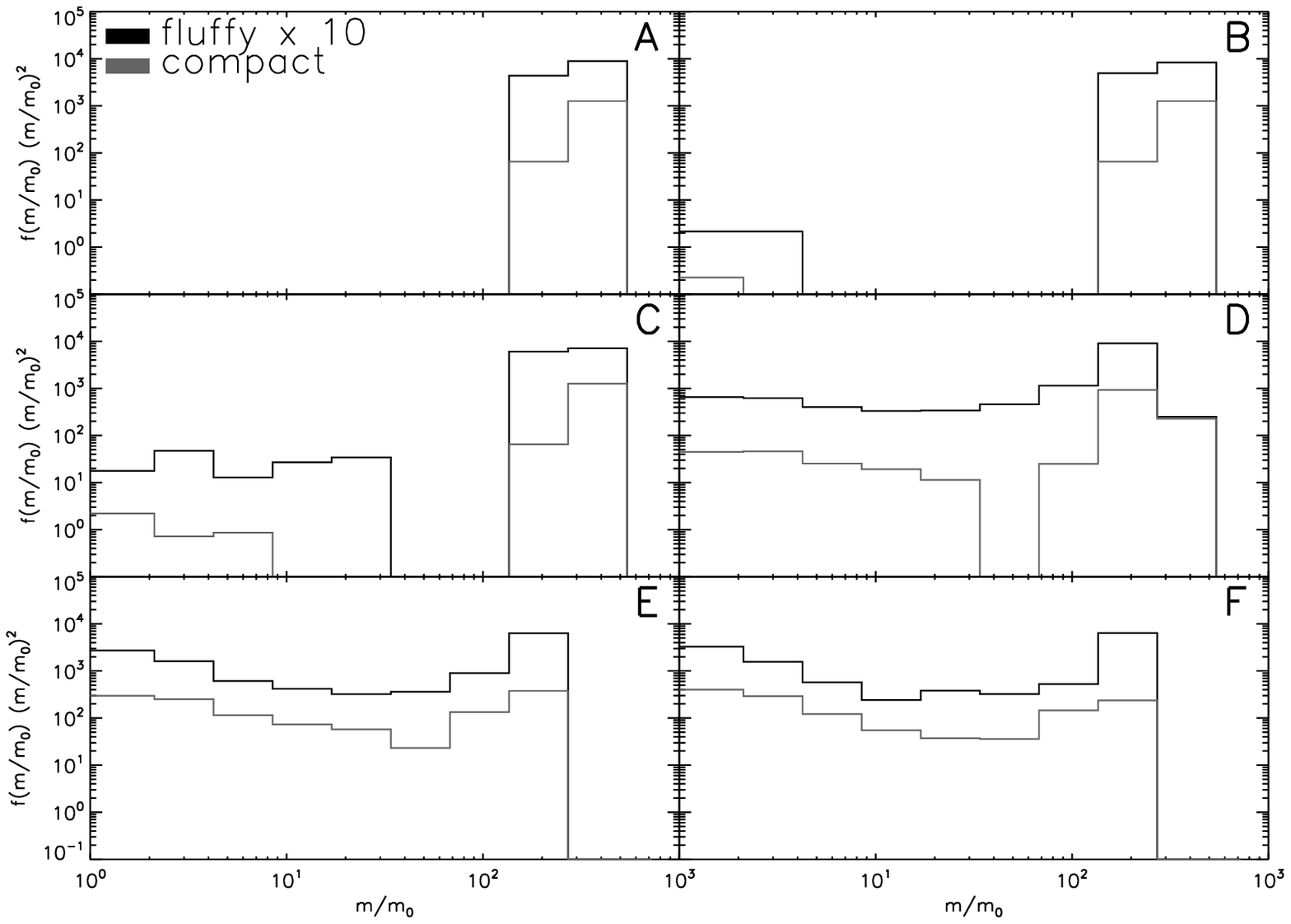}{Distributions averaged over the
  impact parameter. Results for fluffy and compact aggregates at energies A: $E=
  5.7\ep{-4} N E_\mr{br}$, B: $E= 5.7\ep{-2} N E_\mr{br}$, C: $E= 0.13 N
  E_\mr{br}$, D: $E= 0.92 N E_\mr{br}$, E: $E= 8.2 N E_\mr{br}$, and F: $E= 14.6
  N E_\mr{br}$}{distro-baver}
distributions are shown in \fg{distro-baver}.  The panels show the
evolution of mass spectrum for both compact and fluffy aggregates.
The effect of impact averaging is immediately visible through the
presence of the large fragment component at all energies.  At low
energies, growth dominates, and from central collisions one would
expect to only see the large mass component at the total mass.
However, grazing collisions contribute in the original, pre-impact
mass bin, and reduce the average mass of the large fragment component.
An increase in the energy leads to the onset of erosion. Small
fragments appear with a very flat slope of the power-law
distribution. Further increase in the energy enhances the erosion. The
component of small fragments grows and begins to steepen the powerlaw
slope slightly.  Eventually, growth changes into fragmentation. The
large fragment component, however, remains in the distribution.  Using
impact-parameter-averages reduces both the effect of fragmentation and
differences between fluffy and compact particles over what would be
expected from head-on collisions.

\subsection{Compactness evolution}
The filling factor \tpsig, similarly to mass spectrum, shows a different
behavior \cd{in central and offset collisions, respectively.} The
effect of impact parameter averaging is presented in
\fg{psig-baver}. Compact aggregates are decompressed.
\figsmall{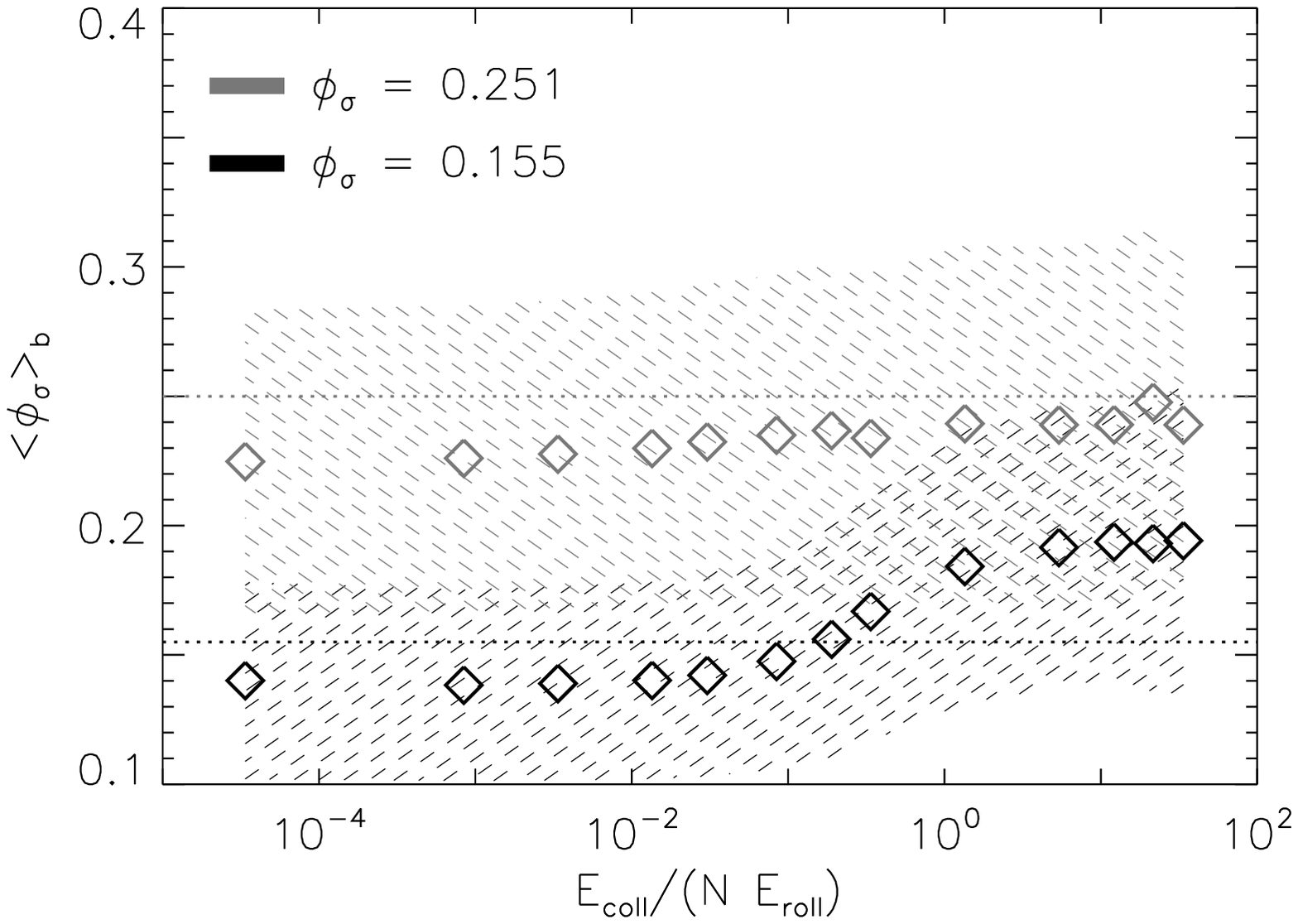}{The filling factor $\phi_\sigma$ averaged
  over the impact parameter vs the collision energy scaled with the number of
  monomers $N$ and in units of the rolling energy $E_\mathrm{roll}$. The dashed
  areas correspond to the spread around the mean. The dotted lines indicate the
  initial value of the filling factor for the colliding particles.}{psig-baver}
At higher energies of about $E=N E_\mr{roll}$, aggregates undergo compression
that mostly compensates the offset collisions and the \emph{dimer-like}
structure discussed in \se{results-ff}.

Fluffy aggregates also increase porosity in low energy collisions.
The increasing energy, however, causes both a decrease in the mass of
the largest fragment and compaction of the aggregates.  \cd{Grazing}
collisions at a high energy produce aggregates with \cd{only} weakly
changed structure.  Therefore, the structural change is dominated by
compressing head-on impacts.  A further increase in the collision
energy results in stronger compression.  The maximum filling factor of
about $\psig\approx 0.19$ is reached at $E=N E_\mr{roll}$.  \cd{Any}
further increase in the energy does not affect the porosity.  Particles
are completely disrupted and the filling factor is dominated by small,
fluffy fragments.

Small fragments produced by erosion or fragmentation are easily
described by a single power-law. Regardless of initial porosity of the
impact energy, small particles have the filling factor given by
\eq{psig-small}.

\subsection{\labelH{sec:global-local}Format}
Our recipe provides parameters required to reconstruct mass
distributions averaged over impact parameter. The \cd{distribution of
  fragment masses} is given by $F(m)\ \mr{d}\!\log m= f(m)\ m^2\
\mr{d}\!\log m$, where $f(m)\ \mr{d}m$ provides the number of
particles of mass $m$ in mass interval between $m$ and $m+\mr{d}m$.
Thus, the functional form of our recipe consists of the two components
and is given by
\begin{equation}
F(m)=\xi_1 m^{q} + \frac{\xi_2 }{\sqrt{2 \pi} \sigma_\mr{G}}
\exp{\frac{-(m-M_\mr{G})^2} {2 \sigma_\mr{G}^2}}, \labelH{eq:fit-function}
\end{equation}
where $\xi_1$ and $\xi_2$ are the normalization constants of the
power-law component and the Gaussian, respectively.  \cd{These
  constants are} not provided since the distribution should be
re-normalized in order to conserve the mass.  Instead we give the
\emph{mass ratio} of the power-law and Gaussian components, which
should be used to determine the mass and normalization constant in
each component.  The power-law component extends from a monomer mass to
a quarter of the total mass.

Our recipe is provided in tabulated form.  The parameters required to
reproduce the collisional outcome may be interpolated linearly.  Our
parameter space is covered very well and spans from very fluffy
aggregates of fractal dimension $D_\mathrm{f}=1.5$ through fluffy
fractal ($D_\mathrm{f}=2.0$) and non fractal PCA aggregates to very
compact particles of $\psig=0.251$.  The energy space is also well
sampled.  Our recipe is based on simulations from hit-and-stick regime
up to a catastrophic destruction.

The main difference between the local and the global sub-recipes stems
from the mass \cd{ratio} between impactor and target.  However, small
projectiles, when carrying sufficient energy (\ie impacting with very
high speeds), may also shatter an entire large target aggregate,
i.e. causing a global effect.  Therefore, the global recipe must be
used not only at mass ratio close to unity, but also at impact
energies sufficient to \emph{globally} affect \cd{the entire}
aggregate. \Fg{algorithm-globloc} presents an algorithm used to
distinguish between the local and the global
\figsmall{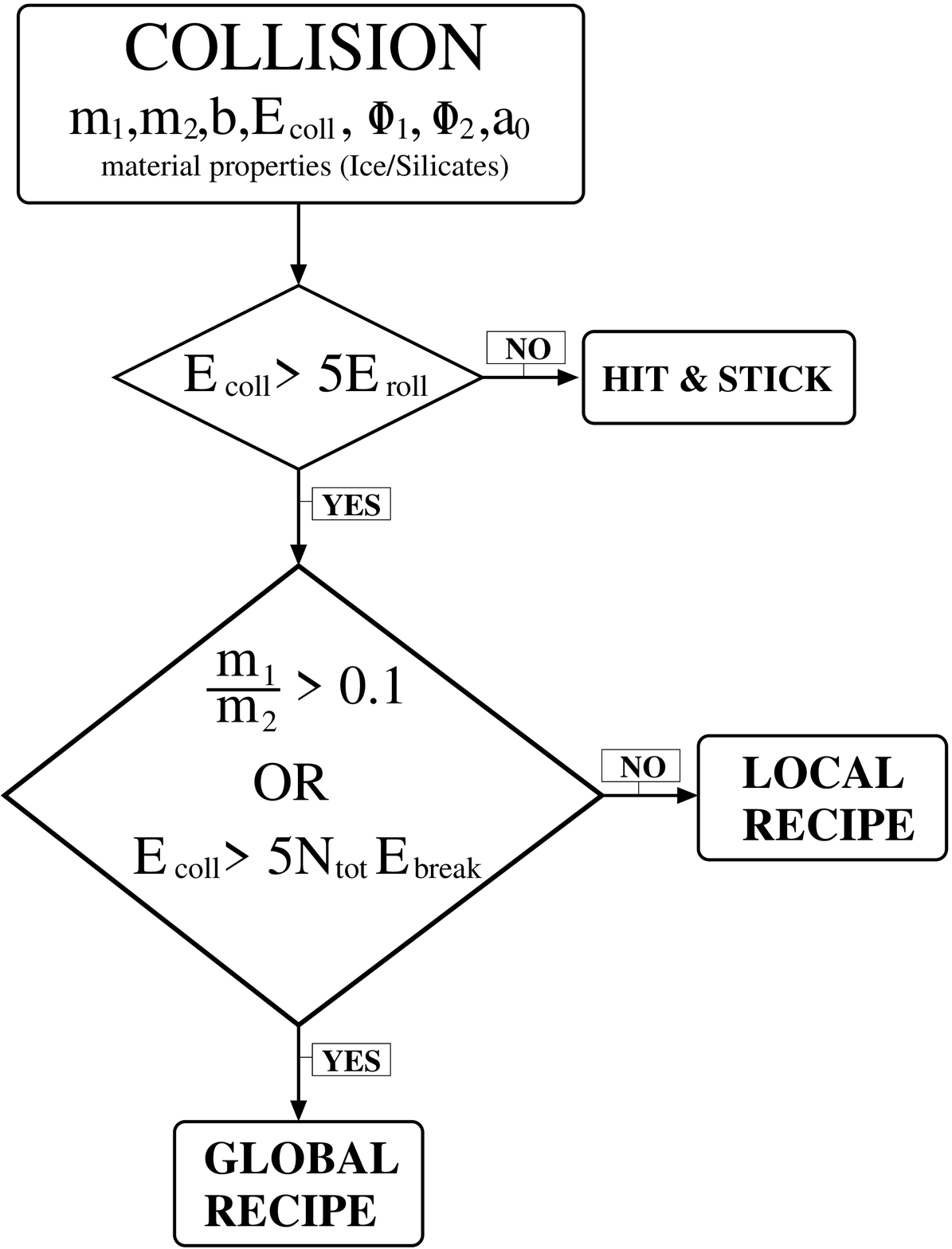}{Algorithm to choose between
  local and global recipes.}{algorithm-globloc}
recipes. When a collision of aggregates characterized by their masses
$m_i$, filling factors $\phi_i$, and some material properties, occurs
at a given impact energy, one must check whether the collision is in
the recipe domain (restructuring or fragmentation) or in the
hit-and-stick regime.  The test checks whether the energy is
sufficient to cause any restructuring.  If the condition
$E_\mathrm{coll} < 5 E_\mathrm{roll}$ is satisfied, the hit-and-stick
mechanism is applied (for details see \citet{2007A&A...461..215O},
\citet{2008OrmelMolcloud_inprep}). Otherwise, a second check is
performed.  The global recipe is applied if any of the two following
conditions is true:
\begin{equation}
\frac{m_1}{m_2} > 0.1,
\labelH{eq:condition-global-mass}
\end{equation}
or
\begin{equation}
E > 5 N E_\mr{br}.
\labelH{eq:condition-global-energy}
\end{equation}
If none of the above is true, the local recipe must be used.

The recipe provides a description of the outcome of a collision between
aggregates of the same filling factor. To describe collision of aggregates with
different compactness, a mass weighted average filling factor must be
determined
\begin{equation}
<\psig>_m = \frac{\sum_i \phi_{\sigma,i} m_i }{\sum_i m_i}.
\labelH{eq:psig-mw}
\end{equation}
Therefore, the filling factor that should be used is dominated by that
of the more massive aggregate.  In particular, the outcome of a
collision between particles of different masses depends very much on the
porosity of the target aggregates and how deep it can be penetrated.
Collisions of equal mass particles, on the other hand, are not
dominated by one species.  Therefore, the contribution from both
aggregates should be about equal.  The compression or decompression is
expected to be weaker than for fluffy particles, as compact aggregates
are more resistant to restructuring (see \se{results}).

%
%
%
%
\subsection{\labelH{sec:rec-tables} A complete quantitative description}
The mass distribution of the collisional outcome for equal mass
colliding projectiles can be constructed using parameters read from
\tb{mean-mass-equal}-\tb{ratio-mass-equal}. \fg{parameters} shows
contour plots
\figlarge{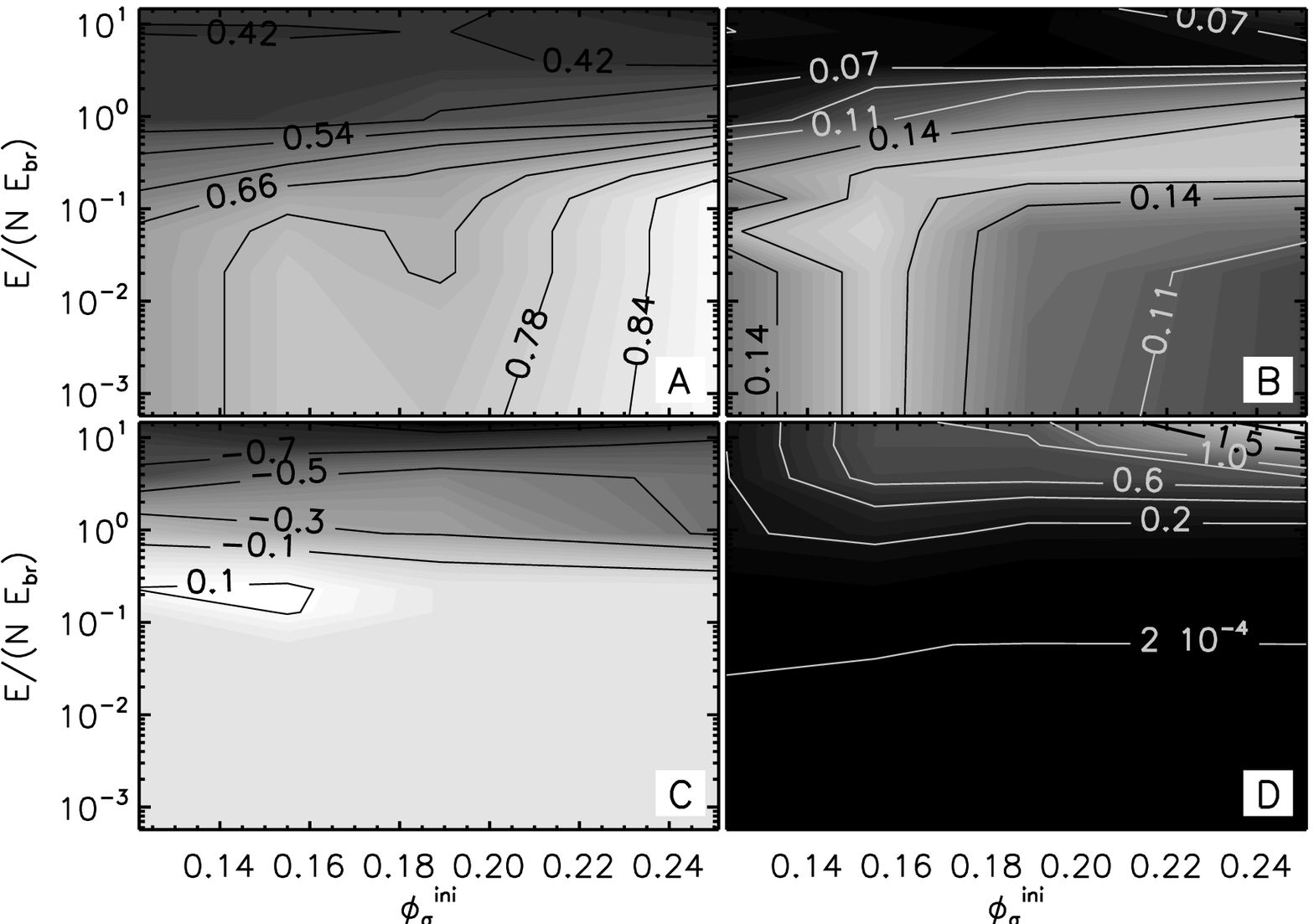}{Parameters as a function of the
  filling factor and impact energy.  Clockwise: mean mass of the
  Gaussian component, width of the Gaussian component, mass ratio of
  the power-law to Gaussian component, slope of the
  power-law.}{parameters}
of the required 4 parameters.  Intuitively, the mean mass of the large
component (upper left panel) decreases with increasing
energy. Similarly, the width of this component (upper right panel)
decreases with increasing energy as a result of fragmentation and
grazing collisions.  Therefore, faster impacts cause formation of the
largest fragments with the lower mass, and the tail of the Gaussian
component decreases.  This shattered mass shifts then to the small
fragments power-law component.  The power-law component does not exist
at low energies and only large aggregates are produced in this stage
due to growth and grazing collisions.  Thus, the slope of this
component (lower left panel) starts to decrease only at larger
energies.  Note that for a weak erosion, where the slope of the
power-law cannot be determined we assume without loss of generality
$q=0$.

\fg{av} shows compression of aggregates in the global recipe. This
contour plot
\figsmall{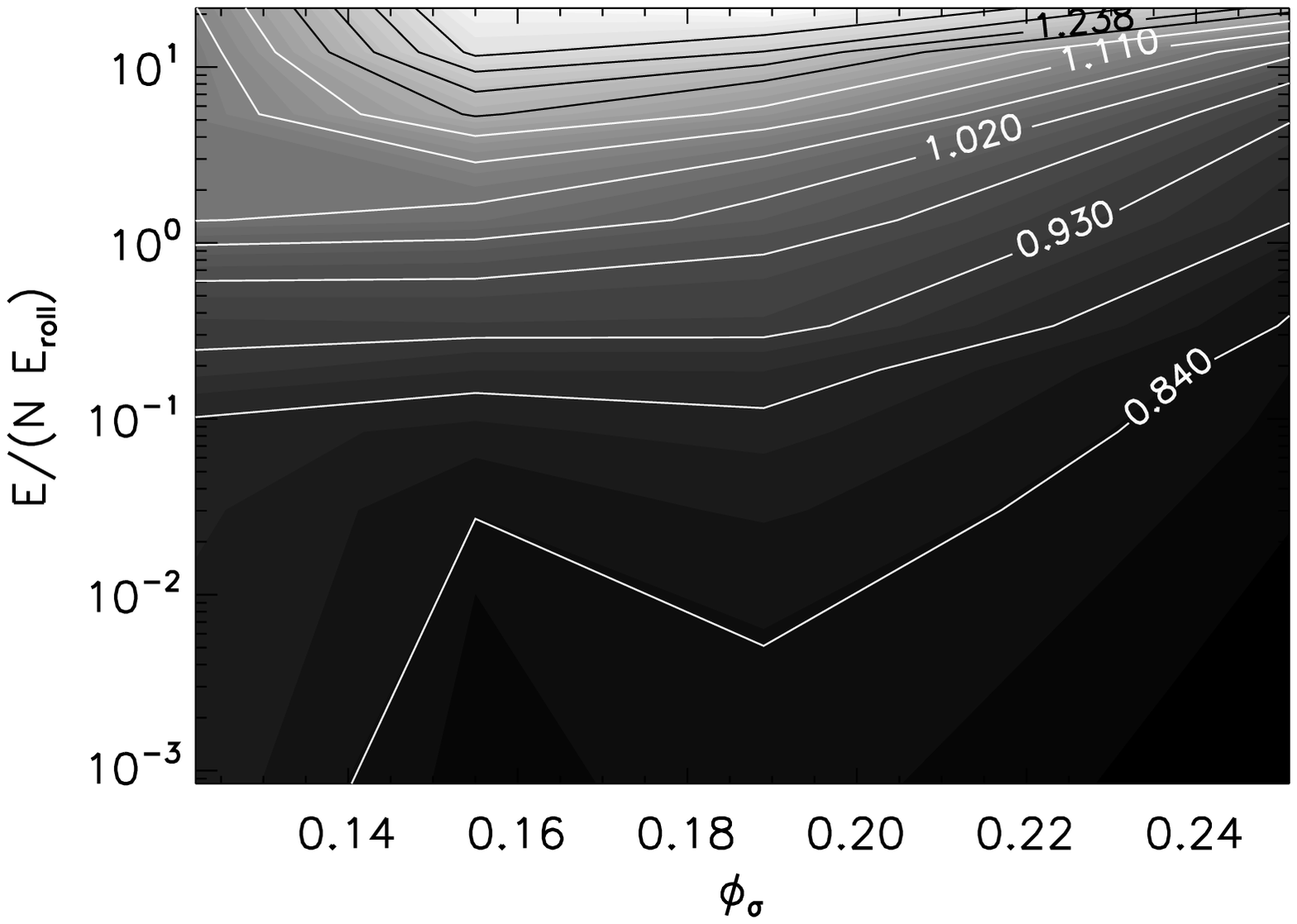}{The change in the filling factor relative to the
  initial filling factor $\psig^\mr{ini}$. $\psig/\psig^\mr{ini}< 1$ indicates
  decompaction and $\psig/\psig^\mr{ini} > 1$ means compaction.}{av}
illustrates \tb{av-compression-equal}. The low-energy collisions cause
a decrease in the filling factor \tpsig\ for all aggregates.  The
effect of decompaction is similar for all particles, although the
strongest decrease in \tpsig\ is observed for the most compact
aggregates.  Faster impacts result in compression and an increase in
the filling factor.  Aggregates with the lowest filling factor \tpsig\
show the lowest structural change.  The density increases only by a
factor of about $\psig/\psig^\mr{ini}=1.11$ at the highest
energies. For very compact aggregates, at the same energy, this change
is somewhat higher ($\psig/\psig^\mr{ini}=1.22$).  This value is
strongly affected by fragmentation.  The average largest collision
remnant in this case consists of only 0.16 of the total mass.  The
maximum compression observed for head-on collisions \cd{at an energy}
of about $E=N\ E_\mr{roll}$ reaches only about 0.89 of the initial
filling factor.  This means that the average largest fragment is
decompressed.  Fluffy aggregates, both fractal and non fractal, show
similar behavior.  For these fluffy particles the boundary between
decompaction and compression is at the energies of about $E=N\
E_\mr{roll}$.  The filling factor increases further at higher
energies.  The maximum compression is reached at highest energies.
The largest fragment at this energy is about a quarter of the total
mass, except of the most fluffy aggregates, that are very difficult to
disrupt completely.

\subsubsection{\labelH{sec:rec-global}Global recipe}
In this section we present tables containing the global recipe. They
describe both the distribution of fragments (\tb{slope-mass-equal},
\tb{ratio-mass-equal}, \tb{mean-mass-equal}, and
\tb{width-mass-equal}) and the change in the filling factor of the
largest fragment, relative to the initial \tpsig\
(\tb{av-compression-equal}).

The shape of the distribution of fragment masses depends on the number
of contacts that can be broken.  The higher the energy is, the
stronger and steeper the power-law component is.  Therefore, the
collision energy is normalized to the breaking energy $E_\mr{br}$.
Moreover, the energy is distributed globally in the aggregates (hence
the name \emph{global recipe}).  Therefore, the energy is also scaled
\cd{with} the total number of particles $N$ (this is approximately
equal to the number of contacts).  The first column in each of
tables~\ref{tab:slope-mass-equal}~--~\ref{tab:width-mass-equal} shows
the normalized energy $E/(N\ E_\mr{br})$.  The following four columns
indicate the recipe output quantity (see \tb{recipe-parameters}) for
aggregates with different initial filling factors
$\psig^\mr{ini}$. \Tb{slope-mass-equal} presents the slope $q$ of the
power-law component of small fragments.  The mass normalization can be
done with the total mass of colliding aggregates $m_1+m_2$ and the
ratio $M_\mr{r}$ of the mass in this power-law component over the
Gaussian component given in \tb{ratio-mass-equal}.

\begin{table}[!hb]
\caption{The slope of the power-law component.}
\centering
\begin{tabular}{r|c|c|c|c}
\hline\hline
$E/(N E_\mr{br})$ & \multicolumn{4}{c}{$\psig^\mr{ini}$} \\
 &  $0.122$ &  $0.155$ &  $0.189$ &  $0.251$ \\
\hline
$5.72\ep{-4}$ & $ 0.00$ & $ 0.00$ & $ 0.00$ & $ 0.00$ \\ 
$2.06\ep{-2}$ & $ 0.00$ & $ 0.00$ & $ 0.00$ & $ 0.00$ \\ 
$5.72\ep{-2}$ & $ 0.00$ & $ 0.00$ & $ 0.00$ & $ 0.00$ \\ 
$1.29\ep{-1}$ & $ 0.00$ & $ 0.11$ & $ 0.00$ & $ 0.00$ \\ 
$2.29\ep{-1}$ & $ 0.10$ & $ 0.12$ & $ 0.00$ & $ 0.00$ \\ 
$9.15\ep{-1}$ & $-0.20$ & $-0.28$ & $-0.31$ & $-0.52$ \\ 
$3.66\ep{0} $ & $-0.69$ & $-0.49$ & $-0.42$ & $-0.53$ \\ 
$8.24\ep{0} $ & $-0.73$ & $-0.81$ & $-0.77$ & $-0.65$ \\ 
$1.46\ep{1} $ & $-0.85$ & $-0.89$ & $-1.03$ & $-0.93$ \\ 
\hline
\end{tabular}
\labelH{tab:slope-mass-equal}
\end{table}

\begin{table}[!hb]
\caption{Mass ratio of the small fragments component (power-law) to the largest
  fragment component (the Gaussian)} 
\centering
\begin{tabular}{r|c|c|c|c}
\hline\hline
$E/(N E_\mr{br})$ & \multicolumn{4}{c}{$\psig^\mr{ini}$} \\
 &  $0.122$ &  $0.155$ &  $0.189$ &  $0.251$ \\
\hline
$5.72\ep{-4}$ & $0.00\cdot10^{-0}$ & $0.00\cdot10^{-0}$ & $0.00\cdot10^{-0}$ & $0.00\cdot10^{-0}$ \\
$2.06\ep{-2}$ & $0.00\cdot10^{-0}$ & $5.43\cdot10^{-5}$ & $0.00\cdot10^{-0}$ & $0.00\cdot10^{-0}$ \\
$5.72\ep{-2}$ & $1.17\cdot10^{-3}$ & $3.26\cdot10^{-4}$ & $8.14\cdot10^{-5}$ & $1.70\cdot10^{-4}$ \\
$1.29\ep{-1}$ & $4.61\cdot10^{-3}$ & $1.47\cdot10^{-2}$ & $4.53\cdot10^{-3}$ & $2.86\cdot10^{-3}$ \\
$2.29\ep{-1}$ & $3.67\cdot10^{-2}$ & $5.66\cdot10^{-2}$ & $1.14\cdot10^{-2}$ & $1.04\cdot10^{-2}$ \\
$9.15\ep{-1}$ & $1.73\cdot10^{-1}$ & $2.67\cdot10^{-1}$ & $1.49\cdot10^{-1}$ & $1.39\cdot10^{-1}$ \\
$3.66\ep{0} $ & $1.91\cdot10^{-1}$ & $6.83\cdot10^{-1}$ & $6.63\cdot10^{-1}$ & $7.87\cdot10^{-1}$ \\
$8.24\ep{0} $ & $2.03\cdot10^{-1}$ & $7.53\cdot10^{-1}$ & $7.57\cdot10^{-1}$ & $1.72\cdot10^{-0}$ \\
$1.46\ep{1} $ & $2.02\cdot10^{-1}$ & $7.45\cdot10^{-1}$ & $8.79\cdot10^{-1}$ & $2.39\cdot10^{-0}$ \\
\hline
\end{tabular}
\labelH{tab:ratio-mass-equal}
\end{table}

The Gaussian component can be reconstructed using three quantities;
the ratio of the mass in the two components $M_\mr{r}$, the mean mass
$M_\mr{G}$, and the width $\sigma_\mr{G}$. \Tb{mean-mass-equal}
provides the mean mass, and the width is given in
\tb{width-mass-equal}.

\begin{table}[!hb]
\caption{Mean mass of the Gaussian component normalized to the total mass.}
\centering
\begin{tabular}{r|c|c|c|c}
\hline\hline
$E/(N E_\mr{br})$ & \multicolumn{4}{c}{$\psig^\mr{ini}$} \\
 &  $0.122$ &  $0.155$ &  $0.189$ &  $0.251$ \\
\hline
$5.72\ep{-4}$ & $0.671$ & $0.756$ & $0.749$ & $0.883$ \\
$2.06\ep{-2}$ & $0.671$ & $0.756$ & $0.711$ & $0.883$ \\
$5.72\ep{-2}$ & $0.671$ & $0.736$ & $0.711$ & $0.883$ \\
$1.29\ep{-1}$ & $0.610$ & $0.697$ & $0.691$ & $0.883$ \\
$2.29\ep{-1}$ & $0.575$ & $0.620$ & $0.671$ & $0.829$ \\
$9.15\ep{-1}$ & $0.430$ & $0.436$ & $0.485$ & $0.536$ \\
$3.66\ep{0} $ & $0.427$ & $0.426$ & $0.423$ & $0.415$ \\
$8.24\ep{0} $ & $0.419$ & $0.418$ & $0.421$ & $0.401$ \\
$1.46\ep{1} $ & $0.422$ & $0.428$ & $0.436$ & $0.359$ \\
\hline
\end{tabular}
\labelH{tab:mean-mass-equal}
\end{table}

\begin{table}[!hb]
\caption{Width of the large fragments component (Gaussian) normalized to the
  total mass.}  
\centering
\begin{tabular}{r|c|c|c|c}
\hline\hline
$E/(N E_\mr{br})$ & \multicolumn{4}{c}{$\psig^\mr{ini}$} \\
 &  $0.122$ &  $0.155$ &  $0.189$ &  $0.251$ \\
\hline
$5.72\ep{-4}$ & $0.124$ & $0.170$ & $0.118$ & $0.098$ \\
$2.06\ep{-2}$ & $0.124$ & $0.170$ & $0.123$ & $0.098$ \\
$5.72\ep{-2}$ & $0.158$ & $0.175$ & $0.123$ & $0.117$ \\
$1.29\ep{-1}$ & $0.120$ & $0.169$ & $0.147$ & $0.137$ \\
$2.29\ep{-1}$ & $0.141$ & $0.164$ & $0.170$ & $0.169$ \\
$9.15\ep{-1}$ & $0.077$ & $0.107$ & $0.135$ & $0.164$ \\
$3.66\ep{0} $ & $0.061$ & $0.065$ & $0.061$ & $0.067$ \\
$8.24\ep{0} $ & $0.070$ & $0.066$ & $0.063$ & $0.071$ \\
$1.46\ep{1} $ & $0.068$ & $0.062$ & $0.061$ & $0.089$ \\
\hline
\end{tabular}
\labelH{tab:width-mass-equal}
\end{table}

The change in porosity of aggregates depends on the number of contacts
that can roll, as the rolling is the main mechanism responsible for
the restructuring.  Therefore, the energy in the first column of
\tb{av-compression-equal} is normalized to the rolling energy
$E_\mr{roll}$. The restructuring also affects the structure of
aggregates globally.  Thus, the collision energy is also normalized to
the total number of monomers, which approximates the initial number of
contacts. \Tb{av-compression-equal} shows the change in the
geometrical filling factor \tpsig\ relative to the initial density
\tpsig$^\mr{ini}$.  This new filling factor represents large fragments
from the Gaussian component.  The power-law component is simply
described by \eq{psig-small}.

The difference in energy scaling applied to the part of the recipe
describing the fragment distribution and to the part of the recipe
providing the compaction, respectively, accounts for cases of
different material properties, where the ratio of the rolling energy
$E_\mr{roll}$ to the breaking energy $E_\mr{br}$ may be different than
for Quartz.  Therefore, different physical processes
(compaction/decompaction and erosion/fragmentation) are \cd{selected}
by different scalings, \ie restructuring scales with the rolling
energy, while the fragmentation scales with the breaking energy.

\begin{table}[!hb]
\caption{A fractional change in the \emph{geometrical} filling factor
  $\psig/\psig^\mr{ini}$, averaged over the impact parameter.}  \centering
\begin{tabular}{r|c|c|c|c}
\hline\hline
$E/(N E_\mr{roll})$ & \multicolumn{4}{c}{$\psig^\mr{ini}$} \\
 &  $0.122$ &  $0.155$ &  $0.189$ &  $0.251$ \\
\hline
$8.44\ep{-4}$ & $0.867$ & $0.818$ & $0.837$ & $0.796$ \\
$3.04\ep{-2}$ & $0.874$ & $0.843$ & $0.860$ & $0.816$ \\
$8.44\ep{-2}$ & $0.877$ & $0.867$ & $0.878$ & $0.822$ \\
$1.90\ep{-1}$ & $0.923$ & $0.901$ & $0.902$ & $0.826$ \\
$3.37\ep{-1}$ & $0.942$ & $0.944$ & $0.943$ & $0.838$ \\
$1.35\ep{0} $ & $1.067$ & $1.053$ & $1.005$ & $0.888$ \\
$5.40\ep{0} $ & $1.082$ & $1.206$ & $1.144$ & $0.937$ \\
$1.21\ep{1} $ & $1.088$ & $1.322$ & $1.274$ & $1.032$ \\
$2.16\ep{1} $ & $1.109$ & $1.368$ & $1.395$ & $1.223$ \\
\hline
\end{tabular}
\labelH{tab:av-compression-equal}
\end{table}


\subsubsection{\labelH{sec:rec-local}Local recipe}
The local recipe describes the outcome of a collision between a small
aggregate and a large target.  \cd{For high} energy impacts, erosion
occurs, resulting in what is basically the large target aggregate
accompanied by a distribution of small fragments.  The energy is
locally distributed over monomers of the small impactor and \cd{some}
surface grains of the target.  Thus, in this case we scale the energy
with the reduced number of monomers $N_\mu = N_1N_2 / (N_1+N_2)$,
which basically is the number of grains in the small aggregate.  As
the erosion is determined by the number of contacts that can be
broken, we scale the collision energy to the breaking energy
$E_\mr{br}$.

\begin{table}[!hb]
\caption{Mean ejected mass relative to the projectile mass. The mass ratio of
  the colliding particles is 0.001.}  \centering
\begin{tabular}{r|c|c|c|c}
\hline
$E/(N_\mu E_\mr{br})$ & \multicolumn{4}{c}{$\psig^\mr{ini}$} \\
 &  $0.07$ &  $0.09$ &  $0.13$ &  $0.16$ \\
\hline
$0.23$  &    $0.97$  &   $0.44$   & $0.52$  &  $0.11$ \\
$0.92$  &    $0.98$  &   $0.58$   & $0.52$  &  $0.11$ \\
$3.66$  &    $1.15$  &   $97.6$   & $3.42$  &  $0.74$ \\
$14.6$  &    $2.70$  &   $109.3$  & $21.3$  &  $7.83$ \\
$33.0$  &    $1.21$  &   $125.1$  & $30.9$  &  $14.7$ \\
$58.6$  &    $2.76$  &   $148.0$  & $34.8$  &  $26.9$ \\
\hline
\end{tabular}
\labelH{tab:mass-different}
\end{table}

\Tb{mass-different} shows the mean mass that is ejected during a
collision, relative to the mass of a smaller aggregate.  The first
column indicates the scaled energy, and the following four columns
show the mean ejected mass for collisions of aggregates of different
initial filling factors \tpsig$^\mr{ini}$.  The mass of the cratered
target aggregate can then be immediately calculated.

The structure modification in the local recipe applies to the large
target aggregate only, as the filling factor of small fragments
follows a simple relation (see \eq{psig-small}).  Bombarding large
aggregates with small projectiles results in very small relative
change in the filling factor (that of the target particle), as the
filling factor quantifies the global structure of the large target
aggregate.  In this case, the energy scaling should thus be done in
respect to the total number of monomers $N$, \cd{which is almost equal
  to} the number of particles in the larger aggregate.  Moreover, the
restructuring mechanisms is determined by the number of monomers that
can roll. Thus the energy is scaled by the rolling energy
$E_\mr{roll}$.

\Tb{av-different} presents the relative change in the geometrical
filling factor for the target particle. The energy listed in the first
column is normalized to the rolling energy $E_\mr{roll}$ and to the
total number of monomers $N$.

\begin{table}[!hb]
\caption{Relative change in the filling factor $\psig/\psig^\mr{ini}$. Mass
  ratio of the two colliding particle is $10^{-3}$.}  
\centering
\begin{tabular}{r|c|c|c|c}
\hline
$E/(N\ E_\mr{roll})$ & \multicolumn{4}{c}{$\psig^\mr{ini}$} \\
 &  $0.07$ &  $0.09$ &  $0.13$ &  $0.16$ \\
\hline
$0.00034$  &  $1.00004$  &  $1.00001$  &  $1.00005$  &  $1.00031$ \\
$0.00135$  &  $1.00002$  &  $1.00011$  &  $1.00964$  &  $1.00072$ \\
$0.00539$  &  $1.00010$  &  $1.04656$  &  $1.00983$  &  $1.00051$ \\
$0.02155$  &  $1.00077$  &  $1.05293$  &  $1.01031$  &  $0.99677$ \\
$0.04849$  &  $1.00012$  &  $1.06677$  &  $1.01024$  &  $0.99314$ \\
$0.08620$  &  $1.00082$  &  $1.06368$  &  $1.00689$  &  $0.98781$ \\
\hline
\end{tabular}
\labelH{tab:av-different}
\end{table}

%
%
%
%
\section{\labelH{sec:conclusions}Conclusions and future work}

In this work we present results of the extensive parameter study of
collisions of three-dimensional aggregates.  The outcome of a
collision is provided in terms of the mass distribution of fragments
as well as the structure of the produced particles.  These simulations
agree with the experimental results in this size regime and provide
scaling that allows for extrapolation to slightly larger sizes as well
as different material properties of monomers (\ie composition and size
of individual grains).

Our simulations indicate new important mechanisms that influence both the
structure and the mass distribution of aggregates:
\begin{itemize}
\item{} The restructuring of aggregates depends mainly on the collision
  energy.  The compaction is reserved for head-on impacts, while offset
  collisions produce elongated and decompacted particles. 
\item{} In the case of erosion and shattering, the structure of small
  fragments \cd{can be written} in a very simple form (see
  \eq{psig-small}), regardless of the initial compactness of colliding
  aggregates or the impact energy.
\item{} The mass distribution of particles produced in a collision
  consists of two individual components. The power-law distribution of
  small fragments is accompanied by a strongly pronounced component of
  large fragments.
\item{} The shape of the mass distribution for the collision outcome
  averaged over the impact parameter is generally independent of the
  impact energy for slow collisions.  At high energies, however, the
  shape of the distribution is almost independent of the structure of
  the colliding particles.
\end{itemize}
These points are combined together in a form of the recipe that
provides the quantitative outcome of a collision.  This recipe is
formulated in a simple way that can be easily applied to models of
dust coagulation in various environments (\eg molecular clouds or
protoplanetary disks). In fact, this recipe has already been applied
to study growth of small dust aggregates in molecular clouds
\citep{2008OrmelMolcloud_inprep}.

\section{Acknowledgment}
We thank J. Blum for useful discussions and hospitality during several
visits.  We also thank M. Min for providing us with the program to
produce the fractal aggregates and C. Ormel and A. Tielens for useful
discussions that significantly improved the final shape of the recipe.
The anonymous referee's careful review has lead to a useful
resturcturing and compaction of the paper.  We acknowledge SARA super
computer center for access to the Lisa computer cluster, which made
this parameter study possible.  We also acknowledge financial support
of Leids Kerkhoven-Bosscha Fonds.  This work was supported by the
Nederlandse Organisatie voor Wetenschapelijk Onderzoek, Grant
614.000.309.

\bibliographystyle{aa}

\end{document}